%% file: main.tex
\documentclass[fleqn,usenatbib]{mnras}

\usepackage[T1]{fontenc}

\DeclareRobustCommand{\VAN}[3]{#2}
\let\VANthebibliography\thebibliography
\def\thebibliography{\DeclareRobustCommand{\VAN}[3]{##3}\VANthebibliography}

\usepackage{graphicx}	
\usepackage{amsmath}	
\usepackage{dblfloatfix}

\usepackage{bm}
\usepackage{amsmath,amsfonts,amssymb,amsthm,gensymb}
\usepackage{array,multirow,makecell} 
\usepackage{xcolor}
\usepackage{newtxtext,newtxmath}
\usepackage{calc,accents}
\newcommand{\dbtilde}[1]{\accentset{\approx}{#1}}

\title[Exact window functions of the 21cm power spectrum]{Impact of instrument and data characteristics in the interferometric reconstruction of the 21\,cm power spectrum}

\input{author-list}

\date{Accepted XXX. Received YYY; in original form ZZZ}

\pubyear{2022}

\begin{document}
\label{firstpage}
\pagerange{\pageref{firstpage}--\pageref{lastpage}}
\maketitle

\begin{abstract}

Combining the visibilities measured by an interferometer to form a cosmological power spectrum is a complicated process. In a delay-based analysis, the mapping between instrumental and cosmological space is not a one-to-one relation. Instead, neighbouring modes contribute to the power measured at one point, with their respective contributions encoded in the window functions. To better understand the power measured by an interferometer, we assess the impact of instrument characteristics and analysis choices on these window functions.
Focusing on the Hydrogen Epoch of Reionization Array (HERA) as a case study, we find that long-baseline observations correspond to enhanced low-$k$ tails of the window functions, which facilitate foreground leakage, whilst an informed choice of bandwidth and frequency taper can reduce said tails.
With simple test cases and realistic simulations, we show that, apart from tracing mode mixing, the window functions help accurately reconstruct the power spectrum estimator of simulated visibilities. The window functions depend strongly on the beam chromaticity, and less on its spatial structure -- a Gaussian approximation, ignoring side lobes, is sufficient. Finally, we investigate the potential of asymmetric window functions, down-weighting the contribution of low-$k$ power to avoid foreground leakage. 
The window functions presented here correspond to the latest HERA upper limits for the full Phase I data. They allow an accurate reconstruction of the power spectrum measured by the instrument and will be used in future analyses to confront theoretical models and data directly in cylindrical space.
\end{abstract}




\section{Introduction} \label{sec:intro}

 As the spin of the electron in the neutral hydrogen atom flips from parallel to anti-parallel, a photon is emitted with the exact wavelength of 21\,cm. Despite this transition being forbidden, the amount of hydrogen present in our Universe (about $75\%$ of all baryonic matter) makes observing the 21\,cm line one of the most exciting prospects of modern cosmology. In the local Universe, most of the signal comes from nearby galaxies, tracing their structure \citep{MartinPapastergis_2010, ObuljenAlonso_2019, HuHoppmann_2019}. In the distant Universe, measuring the 21\,cm signal has the potential of unveiling the reionization of the neutral intergalactic medium (IGM) by the first galaxies on cosmological volumes, for different redshifts \citep[e.g.,][]{CiardiFerrara_2005, FurlanettoOh_2006, MoralesWyithe_2010, PritchardLoeb_2012, Mesinger_2019, LiuShaw_2020}.
 
 Different strategies are used to access the high-redshift \ion{H}{I} signal. Recently, \citet{BowmanRogers_2018} have reported an unexpected absorption profile at 78\,MHz in the sky-averaged spectrum, which has led to numerous interpretations \citep{Barkana_2018, Ewall-WiceChang_2018, MirochaFurlanetto_2019,SinghJishnu_2022}. 
 In order to access more detailed information about the astrophysics of the early galaxies whilst maintaining a good signal-to-noise ratio, another strategy consists of using radio interferometers to measure the power spectrum of the fluctuations of the high-redshift 21\,cm signal. This is the strategy adopted by, e.g., the Giant Metre Wave Radio Telescope \citep[GMRT,][]{Ananthakrishnan_1995} in India, the Low Frequency Array \citep[LOFAR, ][]{vanHaarlemWise_2013} in the Netherlands, the Murchison Widefield Array \citep[MWA,][]{TingayGoeke_2013} in Australia and the Hydrogen Epoch of Reionization Array \citep[HERA, ][]{DeBoerParsons_2017} in South Africa.
 Although none of these experiments has yet achieved a detection, upper limits are getting closer and closer to the cosmological signal \citep{PacigaAlbert_2013, Ewall-WiceDillon_2016, BeardsleyHazelton_2016, ChengParsons_2018, BarryWilensky_2019, LiPober_2019, GehlotMertens_2019, KolopanisJacobs_2019, EastwoodAnderson_2019, TrottJordan_2020, MertensMevius_2020, GarsdenGreenhill_2021, YoshiuraPindor_2021, RahimiPindor_2021}, with HERA reporting the lowest upper limits at $z=7.9$ and $z=10.4$ to date \citep{HERA_upper_limit_H1C_IDR2} with only 18 nights of data and 39 antennas out of the 350 to be deployed. 

There are various reasons why the 21\,cm signal from the Epoch of Reionization has not yet been detected. Notably, the noise levels are still too large for the cosmological signal to be visible in the data. However, this should only be a temporary issue as accumulating data with more observing seasons will increase signal-to-noise ratios (SNR). A more concerning problem is the presence of foregrounds, four to five orders of magnitude brighter than the cosmological signal in the low-frequency range targeted by reionization experiments. Several methods are currently under investigation to either subtract their contribution to the data \citep[e.g.][]{ChapmanAbdalla_2012, ChapmanAbdalla_2013, MertensGhosh_2018, HothiChapman_2021} or simply avoid them, capitalising on their spectral smoothness compared to the cosmological signal. However, the chromaticity of the interferometer's sampling pattern introduces high frequency modulations to the foreground signal, effectively causing them to fill a wedge-like region of the cylindrical space formed by line-of-sight and sky plane Fourier modes, $k_\parallel$ and $k_\perp$, respectively \citep{DattaBowman_2010, MoralesHazelton_2012, VedanthamUdayaShankar_2012, LiuParsons_2014_1, LiuParsons_2014_2, DillonTegmark_2015}. Mathematically, the limit of this wedge is simply set by the delay of a source at the horizon. Outside this wedge, the signal is supposedly dominated by the cosmological signal, forming the `EoR window' for Epoch of Reionization\footnote{Despite the name, the same logic applies to all redshifts, and not only to the EoR.}. Avoiding the wedge is possible: The HERA data are analysed in the `delay approximation' framework, in which the Fourier transform along the frequency axis of a visibility (i.e. a delay transform) is considered analogous to a line-of-sight Fourier transform \citep{ParsonsBacker_2009, ParsonsPober_2012}, effectively concentrating the foregrounds within their wedge. However, imprecise calibration, poor modelling of the beam frequency response, and various other systematics lead to foregrounds leaking from their wedge into the EoR window \citep{DattaBowman_2010, BarryHazelton_2016, Ewall-WiceDillon_2017, OroszDillon_2019, JosephTrott_2020}.

In this paper, we attempt to characterise these effects in order to better understand the power spectrum measured by the interferometer as opposed to the intrinsic cosmological power spectrum. To do so, we derive the expression of the exact window functions of the instrument, which relate the power measured at a given frequency and for a given baseline, to the cosmological power spectrum as a function of $k_\perp$ and $k_\parallel$. The window functions presented here differ from previous works by two main aspects. First, they are obtained outside of the delay approximation and, therefore, fully distinguish between the instrumental baseline-delay $(b, \tau)$\footnote{The delay $\tau$ is the Fourier dual of frequency for a fixed baseline. See Sec.~\ref{subsec:2_wf} for details.} space and the cosmological $(k_\perp, k_\parallel)$ space. Second, they include a precise simulation of the beam along the instrument bandwidth rather than a Gaussian approximation, as well as data weights and flagging.

This paper is organised as follows: In Sec.~\ref{sec:methods}, we present a general derivation of the exact window functions in the framework of a delay-based power spectrum analysis. In Sec.~\ref{sec:data}, we introduce the data and validation simulations used to obtain and test our window functions. In Sec.~\ref{sec:hera}, we present the window functions obtained for the full HERA Phase I data, and assess the impact of various analysis choices and instrument characteristics on the estimated power spectrum. Finally, in Sec.~\ref{sec:3_asym}, we investigate the potential of data analyst-imposed asymmetric window functions to mitigate foreground leakage near the wedge. We discuss our results and conclude in Sec.~\ref{sec:conclusions}. Note that, despite these results being applicable to any low-frequency interferometer, we focus for concreteness on the HERA setup and data, reflective of the latest HERA results obtained with the full Phase I data \citep{HERA_new_upper_limits}.

We use the same \citet{PlanckCollaborationAdam_2016} cosmology as \citet{HERA_new_upper_limits}, with $\Omega_\Lambda = 0.6844$, $\Omega_\mathrm{b} = 0.04911$, $\Omega_\mathrm{c} = 0.26442$, and $H_0 = 67.27\,\mathrm{km/s/Mpc}$.

\section{Methods} \label{sec:methods}

In this section, we first introduce the quadratic estimator formalism used to obtain a cosmological power spectrum from the visibilities measured by the interferometer. We then derive the expression of the exact window functions, outside of the delay approximation.

\subsection{Quadratic estimator formalism} \label{subsec:2_estimator}

In the framework of quadratic estimators of the power spectrum, the continuous quantity $P(\bm{k})$ is discretised by dividing it into bins of predefined thickness in $\bm{k}$-space called the bandpowers. In practice, a bandpower will be built from a set of visibilities measured for a given baseline, on a given frequency band. The estimator $\hat{\bm{p}}$ of the $\alpha^\mathrm{th}$ bandpower is then given by
\begin{equation} \label{eq:estimator_definition}
    \hat{\bm{p}}_\alpha \equiv \bm{x}^\dagger \mathbf{E}^\alpha \bm{x}, 
\end{equation}
where $\bm{x}$ is the data vector -- made of visibilities measured at different frequencies for example, $\mathbf{E}^\alpha$ is a matrix chosen by the data analyst \citep{LiuShaw_2020} and the dagger denotes the Hermitian conjugate. The expectation value of the estimator is then
\begin{equation} \label{eq:estimator_expectation}
\begin{aligned}
\langle \hat{\bm{p}}_\alpha \rangle &= \mathrm{Tr}[\mathbf{E}^\alpha \mathbf{C}]\\ &= \mathrm{Tr}\left[ \mathbf{E}^\alpha \left( \mathbf{C}^{(0)} + \sum_\beta \bm{p}_\beta \mathbf{Q}^\beta \right) \right]\\ &= \sum_\beta \mathrm{Tr}\left[ \mathbf{E}^\alpha \mathbf{Q}^\beta  \right] \bm{p}_\beta + \mathrm{Tr}\left[\mathbf{E}^\alpha \mathbf{C}^{(0)} \right],
\end{aligned}
\end{equation}
where $\mathbf{C}\equiv \langle \bm{x}\bm{x}^\dagger \rangle$ is the data covariance matrix and $\mathrm{Tr}$ stands for the trace of the matrix considered. The covariance depends linearly on the power spectrum \citep{LiuTegmark_2011}, such that
\begin{equation}
    \mathbf{C}=\mathbf{C}^{(0)} + \sum_\alpha \bm{p}_\alpha \mathbf{Q}^\alpha.
\end{equation}
Here, the $\mathbf{C}^{(0)}$ element contains terms that do not depend on the observed sky, such as the instrumental noise covariance. The matrix $\mathbf{Q}^\alpha \equiv \partial \mathbf{C}/\partial \bm{p}_\alpha$ is the response of the covariance matrix to the $\alpha^\mathrm{th}$-bin\footnote{Note that, usually, the matrices $\mathbf{C}$, $\mathbf{E}^\alpha$ and $\mathbf{Q}^\alpha$ are symmetric (Hermitian for complex data).}. In equation~\eqref{eq:estimator_expectation}, the final term is an additive bias term (e.g., an instrumental bias) that comes from the squaring operation inherent to the power spectrum and which vanishes when one correlates two data vectors with different, uncorrelated, noise contributions. Subsequently, we have, for $\bm{p}$ and $\hat{\bm{p}}$ the true and estimated power spectrum, respectively,
\begin{equation} \label{eq:def_W}
    \hat{\bm{p}}= \textbf{W}\bm{p},
\end{equation}
where $\mathbf{W}$ is a matrix such that each row represents a window function and whose elements are given by $\mathbf{W}_{\alpha \beta} \equiv \mathrm{Tr}\left[ \mathbf{E}^\alpha \mathbf{Q}^\beta  \right]$. We call this matrix the window function matrix. Equation~\eqref{eq:def_W} translates the fact that each bandpower estimate is a weighted sum of the true bandpowers. For normalisation purposes, we have, for each bandpower $\alpha$, 
\begin{equation}
\label{eq:norm_wf}
\sum_\beta \mathbf{W}_{\alpha \beta} = 1.
\end{equation}

The matrix $\mathbf{E}^\alpha$ defined in equation~\eqref{eq:estimator_definition} is chosen in order to obtain an optimal estimator, such as a minimal variance estimator. These conditions result in the choice of a normalisation matrix $\textbf{M}$ such that $\hat{\bm{p}} = \textbf{M} \hat{\bm{q}}$ where 
\begin{equation}
    \hat{\bm{q}}_\alpha=\frac{1}{2} \bm{x}_{1}^{\dagger} \mathbf{R}^{\dagger} \mathbf{Q}_{\alpha} \mathbf{R} \bm{x}_{2}
\end{equation}
is the unnormalised estimate of the $\alpha^\mathrm{th}$ bandpower, with $\bm{x}_1$ and $\bm{x}_2$ data vectors, and $\textbf{R}$ a weighting matrix. One can rewrite this equation as $\langle \hat{\bm{q}}\rangle = \mathbf{H}\bm{p}$ 
and identify with equation~\eqref{eq:def_W} to obtain
\begin{equation}
\label{eq:WMH}
    \bf W = MH.
\end{equation}
Choosing $\bf M$ to be diagonal and $\textbf{R}\equiv\textbf{C}^{-1}$ will lead to the minimum variance estimator. Another option is to pick $\mathbf{M}=\mathbf{H}^{-1}$, such that $\langle \hat{\bm{p}}\rangle = \bm{p}$ and the window functions are the identity matrix. However, such a choice artificially inflates the associated error bars on the power spectrum. Finally, one can choose $\bf M$ to diagonalise the covariance of the estimator. In Sec.~\ref{sec:3_asym}, we will see how the normalisation matrix can be modified to obtain desired properties of the window functions such as asymmetry. Note that, in equation~\eqref{eq:def_W}, the matrix $\textbf{W}$ gives the mapping between a baseline-delay $(b, \tau)$ pair and a cosmological $(k_\perp, k_\parallel)$ pair. The simplest form of $\textbf{W}$ is a one-to-one mapping of $b$ to $k_\perp$ and $\tau$ to $k_\parallel$, which is equivalent to making the delay approximation.

\subsection{Delay window functions} \label{subsec:2_wf}

\begin{figure}
    \centering
    \includegraphics[width=.9\columnwidth]{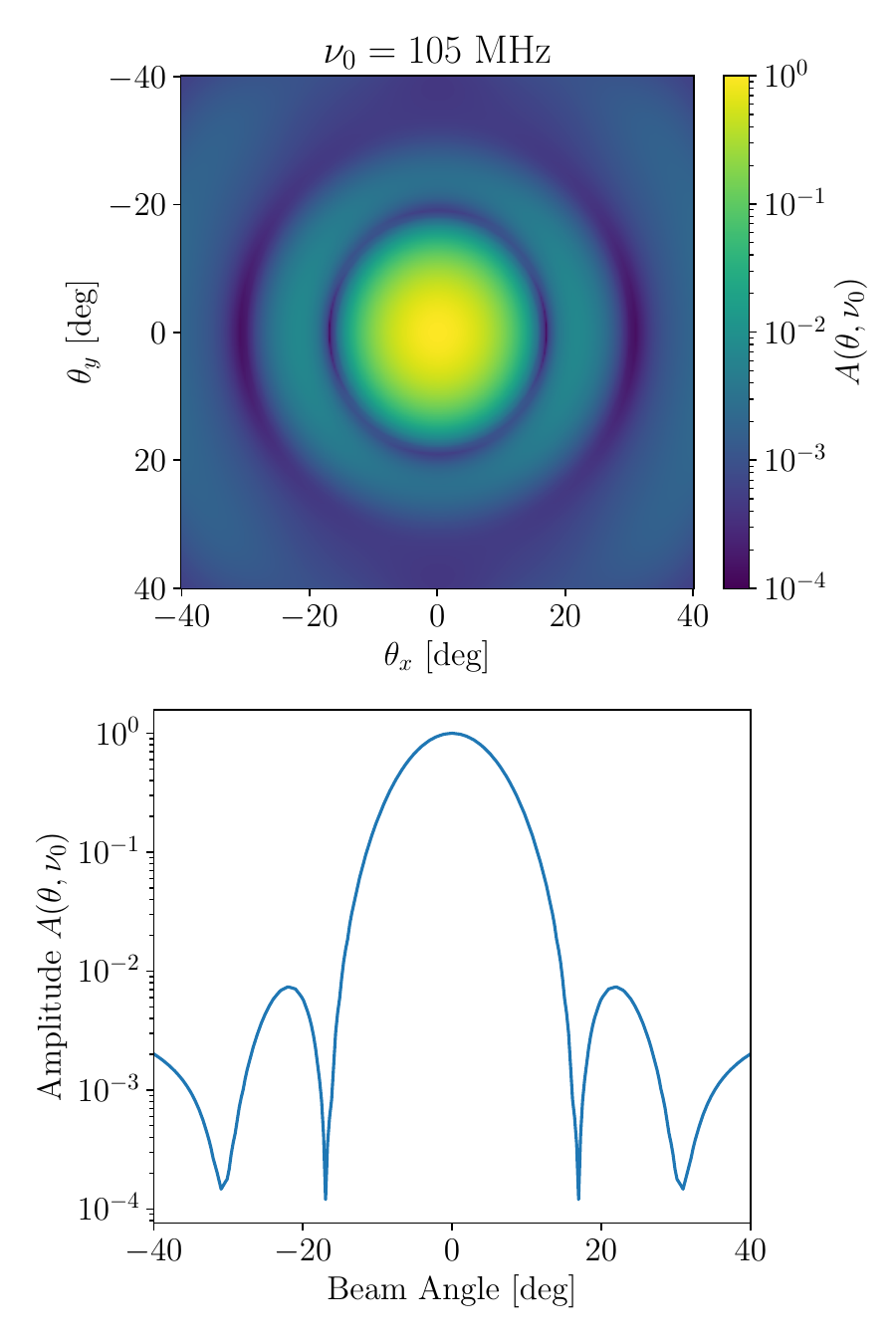}
    \caption{The HERA normalised primary power beam $A(\theta,\nu)$ at $\nu_0=105~\mathrm{MHz}$ for the instrumental \texttt{xx} polarisation, obtained with a simulation \citep{FagnonideLeraAcedo_2021_PhaseI}. This figure illustrates the complicated spatial structure of the beam and the existence of side lobes around the main lobe, centred on the zenith.} 
    \label{fig:HERA_beam}
\end{figure}

\begin{figure*}
    \centering
    \includegraphics[width=.8\textwidth]{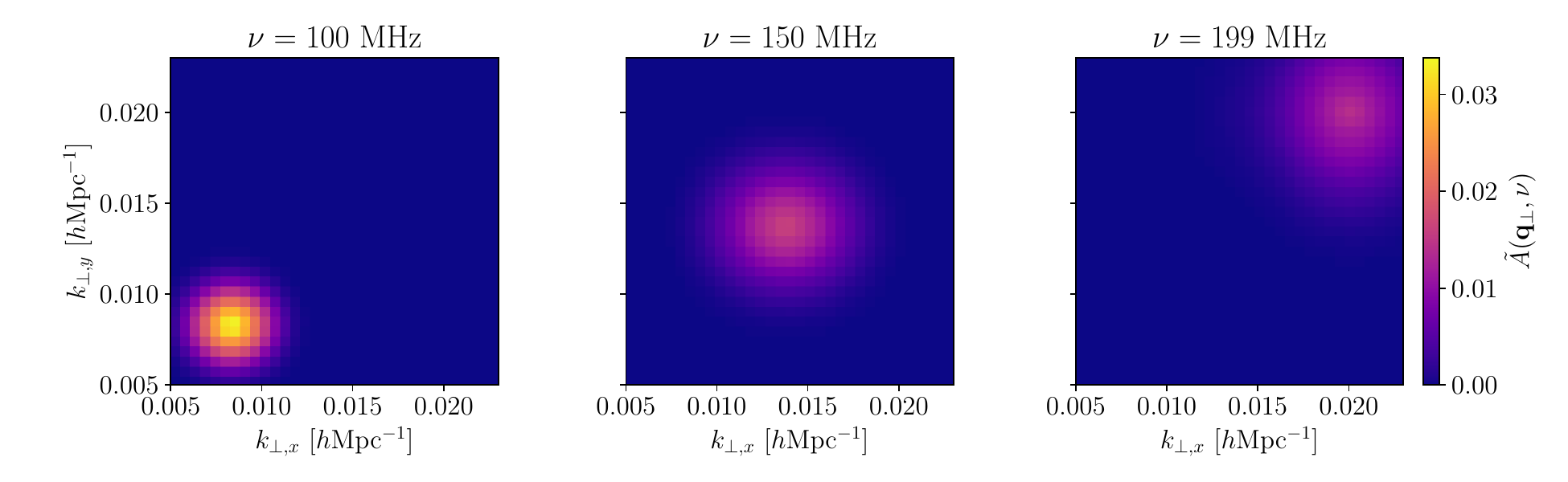}
    \caption{Fourier transform of the HERA beam in the sky plane, represented for different frequencies along the Phase I HERA bandwidth. The shift in Fourier space illustrates the chromaticity of the instrument: At different frequencies, the instrument will probe different spatial scales. The $k_\perp$ coordinates are obtained for $b=38.65\,\mathrm{m}$.}
    \label{fig:Gbeam_shift}
\end{figure*}

We have seen in the previous section that the window functions are both dependent on the instrument, through the matrix $\mathbf{Q}^\alpha$, and of the choice of the data analyst, through the matrix $\mathbf{E}^\alpha$. In this section, we will derive the former contribution, allowing for a better mapping between measurement space ($b$, $\tau$) and cosmological space ($k_\perp$, $k_\parallel$). Here, $b$ is the baseline length and $\tau$ is the delay.
Take the visibility equation
\begin{equation}
\label{eq:visibility_eq}
    V(\bm{b},\nu) = \int \mathrm{d}^2 \bm{\theta} ~ T(\bm{\theta}, \nu) A(\bm{\theta},\nu)\, \mathrm{e}^{-2i\pi \nu \bm{b}\cdot \bm{\theta}/c},
\end{equation}
with $A(\bm{\theta},\nu)$ the primary power beam of the instrument, shown in Fig.~\ref{fig:HERA_beam} for HERA according to the \citet{FagnonideLeraAcedo_2021_PhaseI} simulations, and $T(\bm{\theta}, \nu)$ the sky temperature. In the flat-sky approximation\footnote{\citet{LiuZhang_2016} showed that, despite the large field of view of most 21\,cm experiments, curved-sky corrections to the estimated delay power spectrum are negligible.}, we can re-write the latter in Cartesian coordinates as
$T(\bm{r}_\perp, r_\parallel) \equiv T(\bm{\theta},\nu)$,
with $\bm{r}_\perp\equiv d_c(z) \bm{\theta}$ and $r_\parallel\equiv \alpha(z)\nu $, where $d_c(z)$ is the comoving distance to redshift $z$. We define
\begin{equation}
\alpha(z) \equiv \frac{c(1+z)^2}{\nu_{21}H(z)},
\end{equation} for $\nu_{21}$ the rest-frame 21\,cm frequency, and $H(z)$ the Hubble function.
For a Fourier transform $\tilde{T}(\bm{k}_\perp, k_\parallel)$, one can write
\begin{equation}
T(\bm{r}_\perp,r_\parallel)=\int \frac{\mathrm{d}^2 \bm{k}_\perp \mathrm{d}k_\parallel}{(2\pi)^3} \, \tilde{T}(\bm{k}_\perp, k_\parallel)~ \mathrm{e}^{i (\bm{k}_\perp \cdot \bm{r}_\perp + k_\parallel r_\parallel)},
\end{equation}
which, in turn, leads to
\begin{equation}
\begin{aligned}
V(\mathbf{b},\nu)=\frac{1}{(2\pi)^3} \int \mathrm{d}^2 \bm{\theta}  \int \mathrm{d}^2 & \bm{k}_\perp~ \mathrm{d} k_\parallel\,  A(\bm{\theta},\nu)\, \tilde{T}(\bm{k}_\perp, k_\parallel) \times \\   &\mathrm{e}^{ i \bm{\theta} \cdot [ d_c(z)  \bm{k}_\perp - 2\pi \bm{b}\nu/c]} \mathrm{e}^{i \alpha(z)\nu k_\parallel},
\end{aligned}
\end{equation}
where we have denoted $k_\parallel$ as a scalar since it has a component only along one axis.
We define the delay transform as the Fourier transform of the visibility measured by one baseline, along the frequency axis, according to \citet{ParsonsBacker_2009}:
\begin{equation} \label{eq:delay_transform}
V(\bm{b},\tau) \equiv \int \mathrm{d}\nu\,  V(\bm{b},\nu)~ \mathrm{e}^{-2i\pi \nu \tau} \times \Phi(\nu).
\end{equation}
The visibility inside the integral is multiplied by a tapering function $\Phi(\nu)$ to account for the visibilities being measured on a finite range of frequencies (see Sec.~\ref{subsec:data_hera}). 
Following the equations above, we have
\begin{equation}
V(\bm{b},\tau) =  \frac{1}{(2\pi)^3}  \int \mathrm{d}^2 \bm{k}_\perp~ \mathrm{d} k_\parallel\ \tilde{T}(\bm{k}_\perp, k_\parallel)~ \chi(\bm{k}_\perp,\bm{k}_\parallel;\bm{b},\tau),
\end{equation}
where we have defined the function $\chi$ which describes the mapping between Fourier space and measurement space:
\begin{equation} \label{eq:def_chi}
\begin{aligned}
    \chi(\bm{k}_\perp,k_\parallel;\bm{b},\tau) \equiv \int \mathrm{d}\nu \int \mathrm{d}^2 \bm{\theta}\ & A(\bm{\theta},\nu) \, \mathrm{e}^{ i \bm{\theta} \cdot [ d_c(z)  \bm{k_\perp} - 2\pi \bm{b}\nu/c]} \\ & \times \Phi(\nu)\, \mathrm{e}^{i \nu [ \alpha(z) k_\parallel-2\pi\tau]} .
\end{aligned}
\end{equation}
The estimated delay spectrum can then be written as
\begin{equation}
\hat{P}(\bm{b},\tau) = \frac{1}{(2\pi)^3}  \int \mathrm{d}^3 \bm{k} \ P(\bm{k})\, \vert \chi(\bm{k};\bm{b},\tau) \vert^2,
\end{equation}
where $P(\bm{k})$ is the cosmological, continuous power spectrum.
We see that a bandpower is a weighted sum of the true power spectrum, with the weights being what is usually referred to as the window functions $W(\bm{k};\bm{b},\tau) \propto \vert \chi(\bm{k};\bm{b},\tau) \vert^2$, normalised for each ($\bm{b},\tau$) bin, or each power spectrum estimator, according to
\begin{equation}
\label{ref:eq_wf_norm}
\int \mathrm{d}k_\parallel~\mathrm{d}\bm{k}_\perp~W(\bm{k}_\perp,k_\parallel)=1.
\end{equation}
We then have the continuous equivalent of equation~\eqref{eq:def_W}:
\begin{equation}\label{eq:def_wf}
\hat{P}(\bm{b},\tau) =  \int \mathrm{d}^3 \bm{k} \ P(\bm{k})\,W(\bm{k};\bm{b},\tau).
\end{equation}
Note that these derivations are specific to a delay-spectrum-based analysis and would not carry over to an image-based power spectrum analysis.

Let us now write the full expression giving the window functions. Identifying Fourier transforms in equation~\eqref{eq:def_chi}, we can write
\begin{equation} \label{eq:chi_FT}
\chi(\bm{k};\bm{b},\tau) = \int \mathrm{d}\nu\ \mathrm{e}^{2i\pi \nu [ \alpha(z) k_\parallel/2\pi-\tau]}\,  \tilde{A}(\bm{q}_\perp,\nu)\times \Phi(\nu),
\end{equation}
where $\tilde{A}(\mathbf{q}_\perp,\nu)$ is the Fourier transform of $A(\theta, \nu)$ in the sky plane, with Fourier dual
\begin{equation} \label{eq:def_qperp}
\bm{q}_\perp \equiv\frac{\nu}{c}\bm{b}- \frac{d_c(z)}{2\pi }  \bm{k_\perp}.
\end{equation}
We recognise the commonly-used $\bm{u}$ coordinate defined as $\bm{u}\equiv \nu \bm{b}/c$.
In the delay approximation, the frequency-dependent term vanishes and the Fourier dual of $\bm{\theta}$ is simply $d_c(z) \bm{k_\perp}/2\pi $: We recover the fact that the approximation is valid for short baselines \citep[$b\sim0$, ][]{ParsonsPober_2012}. Outside of this approximation, the chromaticity of the beam translates as a shift by $\nu b / c$ in the Fourier transform of the beam, as illustrated on Fig.~\ref{fig:Gbeam_shift}.

\begin{figure}
    \centering
    \includegraphics[width=\columnwidth]{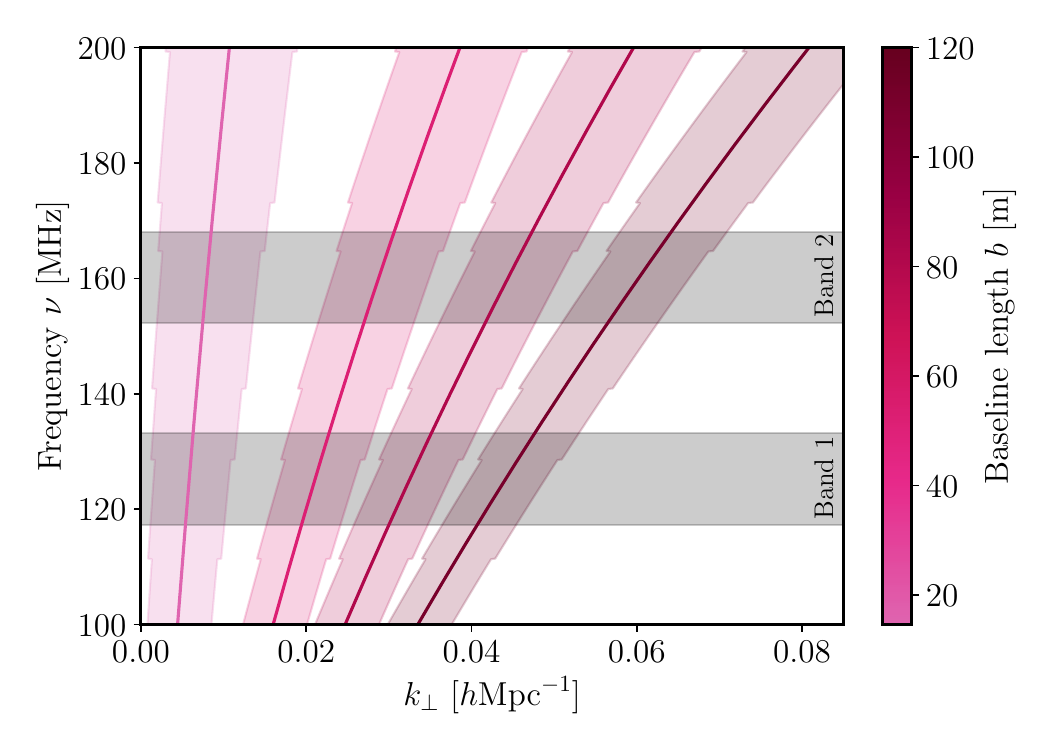}
    \caption{Perpendicular (sky plane) Fourier modes probed by different baselines along the HERA bandwidth, illustrating how the range of $k_\perp$ probed by a given baseline evolves with frequency. Solid lines correspond to equation~\eqref{eq:wf_centres} and shaded areas to $99\%$ of the total integral of the Fourier transform of the beam $\tilde{A}(\bm{q}_\perp, \nu)$ for each frequency. The jagged edges are a result of the limited resolution of the beam simulation. This figure is another illustration of the variety of spatial scales probed by a given baseline along the instrument's bandwidth. In the delay approximation, only the one-to-one mapping represented by the solid lines is considered.}
    \label{fig:freq_vs_kperp}
\end{figure}

The final integral over the frequency can also be considered a Fourier transform, where $\eta$ is the Fourier dual of $\nu$ such that
\begin{equation} \label{eq:def_eta}
\eta \equiv \tau - \frac{\alpha(z)}{2\pi} k_\parallel.
\end{equation}
Again, in the delay approximation, the second term vanishes, and the Fourier dual of the frequency is simply the delay $\tau$. 
We take the Fourier transform of $\tilde{A}(\mathbf{q}_\perp,\nu)$ along the frequency axis to obtain $\tilde{\tilde{A}}$, and write
\begin{equation}
\label{eq:w_Atilde}
W(\bm{k}_\perp,k_\parallel;\bm{b},\tau) = \left\vert \dbtilde{A}\left(\frac{\nu}{c}\bm{b}- \frac{d_c(z)}{2\pi }  \bm{k}_\perp,\tau - \frac{\alpha(z)}{2\pi} k_\parallel\right) \times \tilde{\Phi}(\eta) \right\vert^2,
\end{equation}
which we can cylindrically average to obtain $W(k_\perp,k_\parallel;\bm{b},\tau)$\footnote{Note that these window functions are symmetrical with respect to $\tau$, that is
$W(k_\perp,k_\parallel;\bm{b},\tau)=W(k_\perp,k_\parallel;\bm{b},-\tau)$.}. If the primary beam peaks at zenith, each window function corresponding to a $(b,\tau)$ pair will peak at a $(k_\perp, k_\parallel)$ pair given by:
\begin{equation} \label{eq:wf_centres}
\left\{ \begin{aligned}
&k_\perp =  \frac{2\pi}{d_c(z)} \frac{\nu b}{c} ,  \\
&k_\parallel = \frac{2\pi \vert\tau\vert}{\alpha(z)} .
\end{aligned}\right.
\end{equation}
In Fig.~\ref{fig:freq_vs_kperp}, we illustrate how the $k_\perp$ probed by a given baseline evolves with frequency, according to the equation above. This is not a one-to-one mapping: Effectively, each baseline integrates over a range of $k_\perp$ modes, increasing with frequency and with baseline length. Note that the width of this range is given by the width of the Fourier transform of the beam shown in Fig.~\ref{fig:Gbeam_shift}, clearly frequency-dependent.

For a Gaussian beam, the window functions can be derived analytically, greatly simplifying the computations and avoiding resolution issues.
Indeed, integrating by parts, the Fourier transform $\tilde{A}(\bm{q})$ of a Gaussian beam $A(\bm{\theta})$ defined as a function of the flat-sky angle $\theta$ is: 
\begin{equation}
    A(\bm{\theta})=\mathrm{exp}\left[-\theta^2/ \sigma_b(\nu)^2\right] \longleftrightarrow \tilde{A}(\bm{q}) \propto \mathrm{exp}\left[-\pi q^2 \sigma_b(\nu)^2 \right],
\end{equation}
where $\sigma_b(\nu)$ is the width of the beam, for which we model the frequency and polarisation dependency of the HERA beam. This process is described in more details in Appendix~\ref{app:Gaussian_beam}.

We will apply this formalism to two data sets, depending on the tests we wish to perform: The full HERA Phase I data set or the simulations used to validate the initial Phase I analysis. They are both introduced in Sec.~\ref{sec:data} below.

\section{Data}
\label{sec:data}

To assess the impact of window functions on the estimator of the cosmological power spectrum, we use throughout this paper the HERA data presented in \cite{HERA_new_upper_limits}, as well as the validation simulations introduced in \citet{HERA_validation_2022}. We describe some essential features of both these data sets below, but refer the reader to the aforementioned papers for more details.

\subsection{The HERA Phase I data}
\label{subsec:data_hera}

In contrast to the initial analysis presented in \citet{HERA_upper_limit_H1C_IDR2}, the new data set includes a full season of HERA Phase I data, that is 94\,nights of observing (Julian dates 2458026 to 2458208) spanning a period from September 2017 to March 2018. Depending on the nights and selection criteria applied, up to 42 antennae are used, forming baselines whose lengths range from 14.6\,m to 124.8\,m. The Phase I observing setting is made of HERA's 14-m parabolic antennae with cross-dipole feeds, front-end and back-end systems inherited from the PAPER experiment \citep{ParsonsBacker_2010, ChengParsons_2018, KolopanisJacobs_2019, FagnonideLeraAcedo_2021_PhaseI}. Observations cover frequencies between 100 and 200\,MHz, corresponding to redshifts $6.1 \leq z \leq 13.2$. The full bandwidth is run through the data reduction pipeline, which includes redundant-baseline and absolute calibration, radio frequency interference (RFI) flagging, gain smoothing, LST binning, hand-flagging, inpainting of flagged frequency channels, and cross-talk subtraction  \citep{DeBoerParsons_2017, KernParsons_2020, KernDillon_2020, DillonLee_2020}. However, only two frequency bands are selected for the power spectrum analysis, in order to avoid sections with heavy flagging:
\begin{description}
    \item Band 1: $117.1 \leq \nu / \mathrm{MHz} \leq 132.6$, centred on $z=10.4$, and
    \item Band 2: $150.3 \leq \nu / \mathrm{MHz} \leq 167.8$, centred on $z=7.9$.
\end{description}
Note that these are slightly different from the ones used in the first Phase I analysis \citep{HERA_upper_limit_H1C_IDR2}. In order to avoid edge effects in the Fourier transforms, we apply a Blackman-Harris tapering function along the spectral window of each band \citep{Blackman1958}, effectively reducing their bandwidth by a half. Another important characteristic of the data set is that, in order to to avoid bright foregrounds in the sky, such as the Galaxy or Fornax A, the $10^\circ$-wide stripe centred on declination $-30.7^\circ$ covered by the HERA drift scan is divided into five `fields', corresponding to cuts in the data in local sidereal time (LST). The power spectrum is estimated independently on each of those fields, corresponding to LST ranges of 21.5–0.0\,hours, 0.75–2.75\,hours, 4.0–6.25\,hours, 6.25–9.25\,hours, and 9.25–14.75\,hours.

An additional important analysis choice to highlight here is the wedge buffer. This buffer corresponds to modes within the EoR window excluded from the spherical power spectra estimates, as the beam chromaticity, the usage of a tapering function, and their intrinsic chromaticity lead foregrounds to leak outside of their wedge \citep{ParsonsPober_2012}. In \citet{HERA_new_upper_limits}, we choose this buffer to be $300\,\mathrm{ns}$ away from the horizon wedge ($200\,\mathrm{ns}$ in the previous analysis), corresponding to $k=0.15\,h\mathrm{Mpc}^{-1}$ for Band~1 and $k=0.17\,h\mathrm{Mpc}^{-1}$ for Band~2.

We use the `power spectrum method' of \citet{TanLiu_2021} to estimate the error bars on the measured power spectrum. These rely on the noise power spectrum $P_N$ and on an unbiased estimator of the noise and signal-noise cross-term $\hat{P}_{SN}$. The former is defined by
\begin{equation}
\label{eq:noise_ps}
P_{N}=\frac{\alpha^{2}(z)\, d_c(z) \, \Omega_\mathrm{eff}\, T_\mathrm{sys}^{2}}{t_\mathrm{int} N_\mathrm{co} \sqrt{2 N_{\mathrm{inco}}}},
\end{equation}
where $\Omega_\mathrm{eff}$ is the effective beam area, $T_\mathrm{sys}$ is the system temperature, $t_\mathrm{int}$ is the integration time, and $N_\mathrm{co}$ and $N_\mathrm{inco}$ are, the number of integrations averaged together coherently and incoherently, respectively\footnote{An coherent average is done prior to forming the power spectrum, whilst an incoherent average is performed after.}. The latter writes
\begin{equation}
\hat{P}_{SN}=P_N \times \sqrt{\sqrt{2} \hat{P} / P_N +1} \times \left( \sqrt{1 / \sqrt{\pi}+1}-1 \right).
\end{equation}
We refer the interested reader to \citet{TanLiu_2021} for a more detailed description of the error bar estimation.

\begin{figure}
    \centering
    \includegraphics[width=\columnwidth]{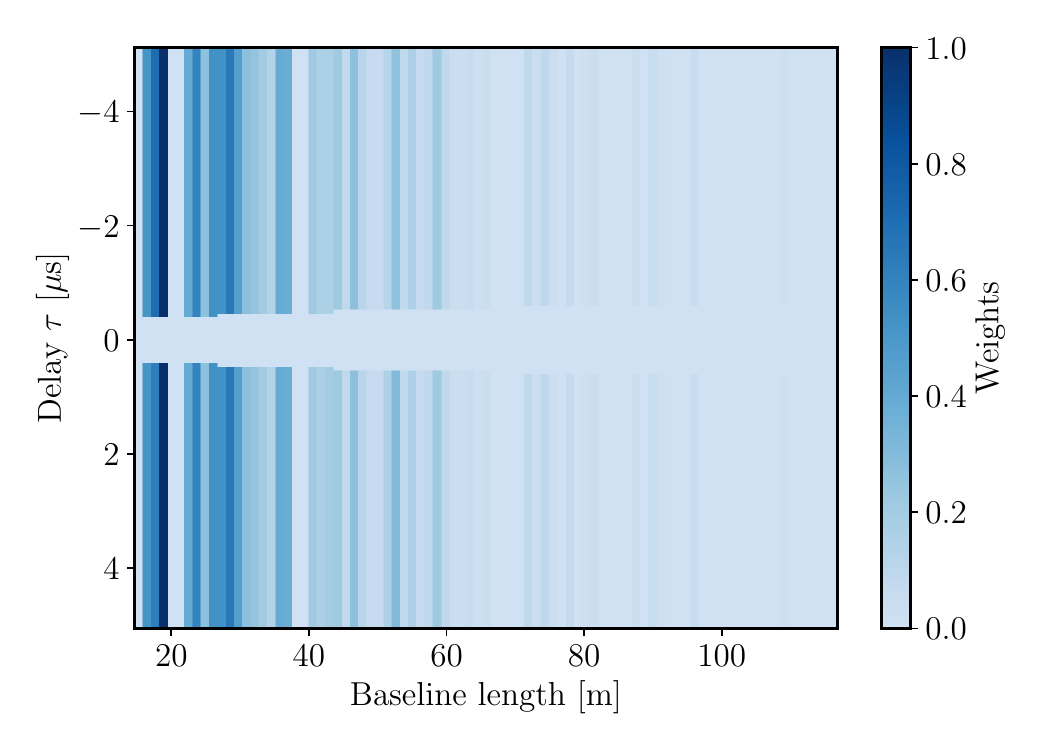}
    \includegraphics[width=\columnwidth]{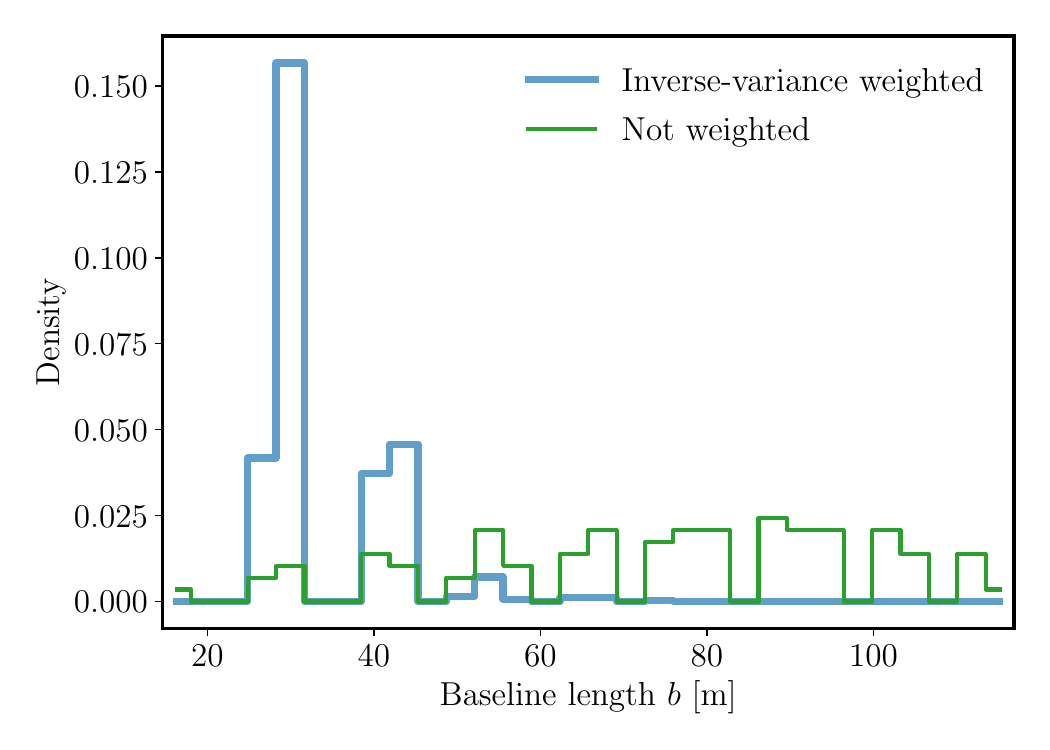}
    \caption{\textit{Upper panel:} Normalised inverse-noise variance weights applied to the full Phase I HERA data along Band~2. A zero weight corresponds to infinite variance. Flagged data, such as data located in the wedge buffer ($\tau \sim 0$), has zero weight. \textit{Lower panel:} Contribution of baselines of different lengths to the HERA Phase I data along Band~2, before (thin green line) and after (thick blue line) applying inverse-variance weights. Because of the high redundancy of the array, the shorter baselines have better sampling and lower noise, explaining their significant contribute to the signal after applying weights. The shortest baselines ($b \lesssim 20\,\mathrm{m}$) are flagged because of cross-talk.}
    \label{fig:hist_weights}
\end{figure}

When incoherently averaging the power spectra over redundant baseline groups and within fields, we apply weights corresponding to the inverse square of the noise power spectrum. This corresponds to an inverse variance-weighted average and the same weights are later applied to the window functions in Sec.~\ref{sec:hera}. Note that these weights also include flagging, with flagged data having zero weight (infinite variance). Overall, about $35\%$ of all individual baseline-delay pairs along Band~2 have infinite variance. The impact of these weights on the contribution of different baselines to the final power spectrum is illustrated on Fig.~\ref{fig:hist_weights}. We see that the noise-variance correction removes the contribution from short baselines, that have been flagged because of cross-talk residuals. Additionally, the small number of long baselines leads to a high noise in their sampling, so that applying inverse-variance weights reduces their contribution to the final power spectrum.

\subsection{Validation simulations}
\label{subsec:data_sims}

To understand the impact of the window functions on EoR signal recovery, we use the end-to-end simulations presented in \citet{HERA_validation_2022}. The simulated clean visibilities include a cosmological signal and foregrounds. The mock foregrounds are made of the spatially-smooth diffuse emission from the Galaxy, obtained with the Global Sky Model \citep[GSM, ][]{GSM}, and point-like sources from the GLEAM catalogue \citep{Hurley-WalkerCallingham_2017, ZhengTegmark_2017, KimLiu_2018} and additional bright sources such as Fornax A \citep{McKinleyYang_2015}. The mock EoR signal is a Gaussian random field with a time-invariant power spectrum such that $P_\mathrm{true}(k,z)=A_0 k^{-2}$, converted to its angular harmonic space analogue $C_\ell(\nu, \nu')$. Corresponding harmonic realisations are produced\footnote{The simulator used is available at \url{https://github.com/zacharymartinot/redshifted gaussian fields}.} and run through \texttt{RIMEz}\footnote{Available at \url{https://github.com/UPennEoR/RIMEz}.} to generate visibilities. The clean visibilities are then contaminated with all instrumental effects known for HERA, including thermal noise, antenna gains, cross-coupling, and cable reflections. Data sets with different components are run through the full Phase I analysis pipeline (see Sec.~\ref{subsec:data_hera}) and the subsequent power spectrum estimation pipeline, \texttt{hera\_pspec}\footnote{Available at \url{https://github.com/HERA-Team/hera_pspec}.}. In this work, we use the power spectra made with EoR signal only, foregrounds only, EoR and foregrounds but no systematics, and EoR, foregrounds, systematics and instrumental effects. 

The main result of \citet{HERA_validation_2022} is that, for all bands and fields considered, the HERA analysis pipeline produces unbiased power spectrum estimates consistent with the known analytic input at the $2\sigma$ level for $k>0.2~h\mathrm{Mpc}^{-1}$, where the EoR signal dominates the foregrounds. On even smaller scales ($k \gtrsim 0.4~h\mathrm{Mpc}^{-1}$), the recovered signal matches the predicted noise floor $P_N$, showing that systematics are mitigated below the noise level.

Note that the simulations used to validate the results presented in \citet{HERA_new_upper_limits} are slightly different from the ones described above. Namely, the EoR signal has a boosted amplitude in the new simulations, which has the advantage of allowing for a more precise estimate of analysis biases, but is not useful for this work. Therefore, we use the validation simulations corresponding to the limited data set used in \citet{HERA_upper_limit_H1C_IDR2}.\\

We now turn to estimating the window functions described in Sec.~\ref{sec:methods}. first in cylindrical space, and then their spherical average, for the two data sets described above.

\section{Results} \label{sec:hera}

In Sec.~\ref{subsec:3_delay}, we apply the formalism derived in Sec.~\ref{sec:methods} to obtain the exact window functions corresponding to the analysis of the full Phase I HERA data \citep{HERA_new_upper_limits} described in Sec.~\ref{subsec:data_hera}. We then study the impact of different analysis choices and instrument characteristics, such as the frequency resolution or the bandwidth, on the resulting window functions in Sec.~\ref{subsec:3_impact_props}. In Sec.~\ref{subsec:3_validation}, we illustrate the importance of knowing the exact window functions of one's estimator with the help of simple test cases and more realistic simulations. Finally, in Sec.~\ref{sec:3_asym}, we investigate the possibility of including asymmetric window functions to the analysis, in the hope of reducing the foreground leakage from low to high $k$-modes.

Throughout this section, we will call `approximate' window functions the window functions obtained in the framework of the delay approximation. On the other hand, the `exact' window functions are the ones obtained following the calculations of Sec.~\ref{subsec:2_wf}.

\begin{figure*}
    \centering
    \includegraphics[width=.85\textwidth]{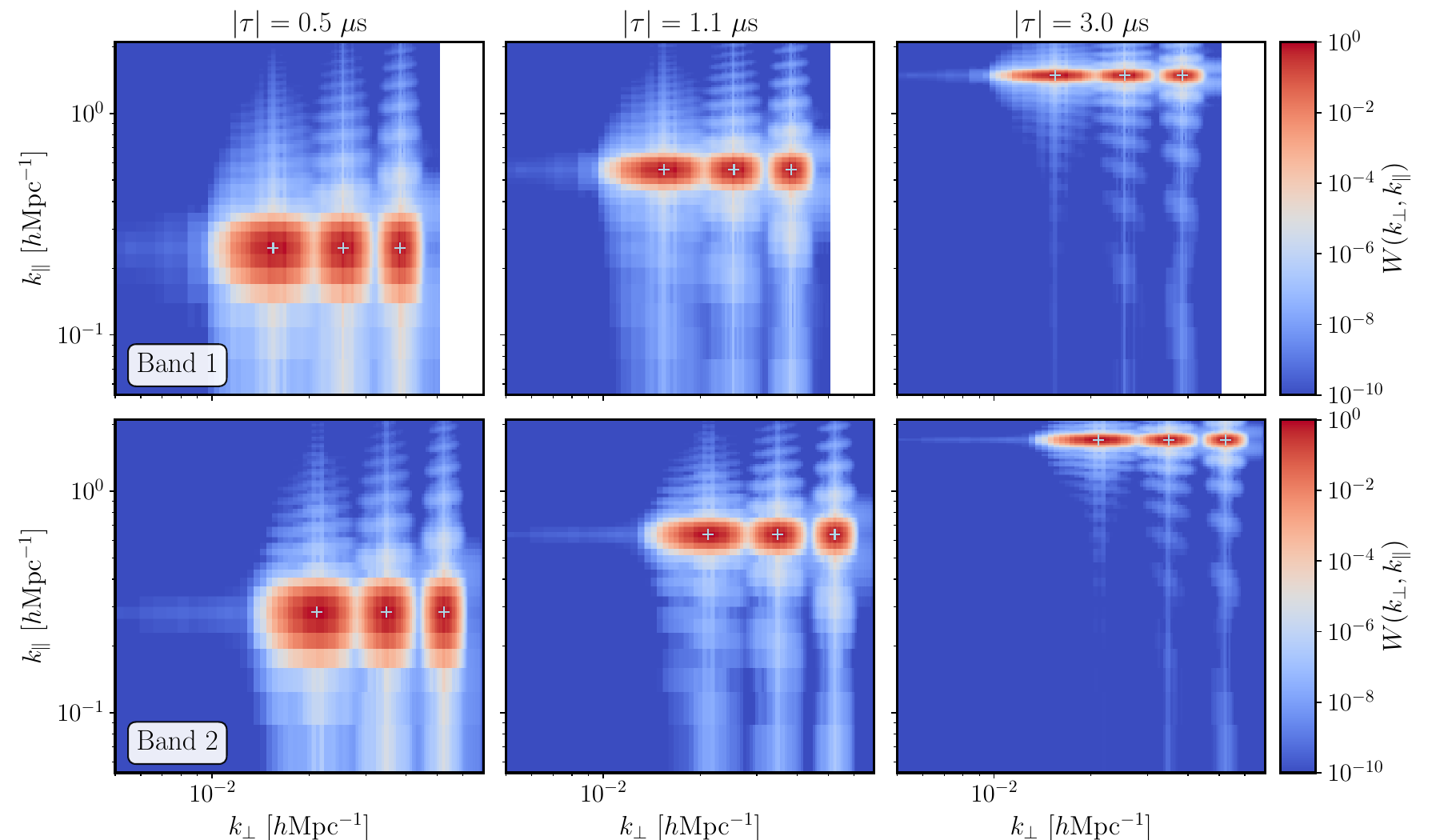}
    \caption{Example of cylindrical window functions $W(b,\tau; k_\perp, k_\parallel)$ obtained with the formalism described in Sec.~\ref{sec:methods} for HERA Band~1 (first row) and Band~2 (second row). Each panel corresponds to a different delay, and, within each panel, the window functions corresponding to baseline lengths 38.7, 63.7, 95.8 meters are represented. The centre (maximum) of each window function in $(k_\perp, k_\parallel)$ space is given by equation~\eqref{eq:wf_centres}. Note that the $k_\parallel$ value corresponding to each delay $\tau$ will slightly change between Band~1 and Band~2. This figure illustrates the contribution of neighbouring modes to the power spectrum measured at a given point (the centre of the window function, identified with equation~\eqref{eq:wf_centres} and shown as a cross).}
    \label{fig:cylindrical_wf_Band1}
\end{figure*}

\begin{figure*}
    \centering
    \includegraphics[width=\textwidth]{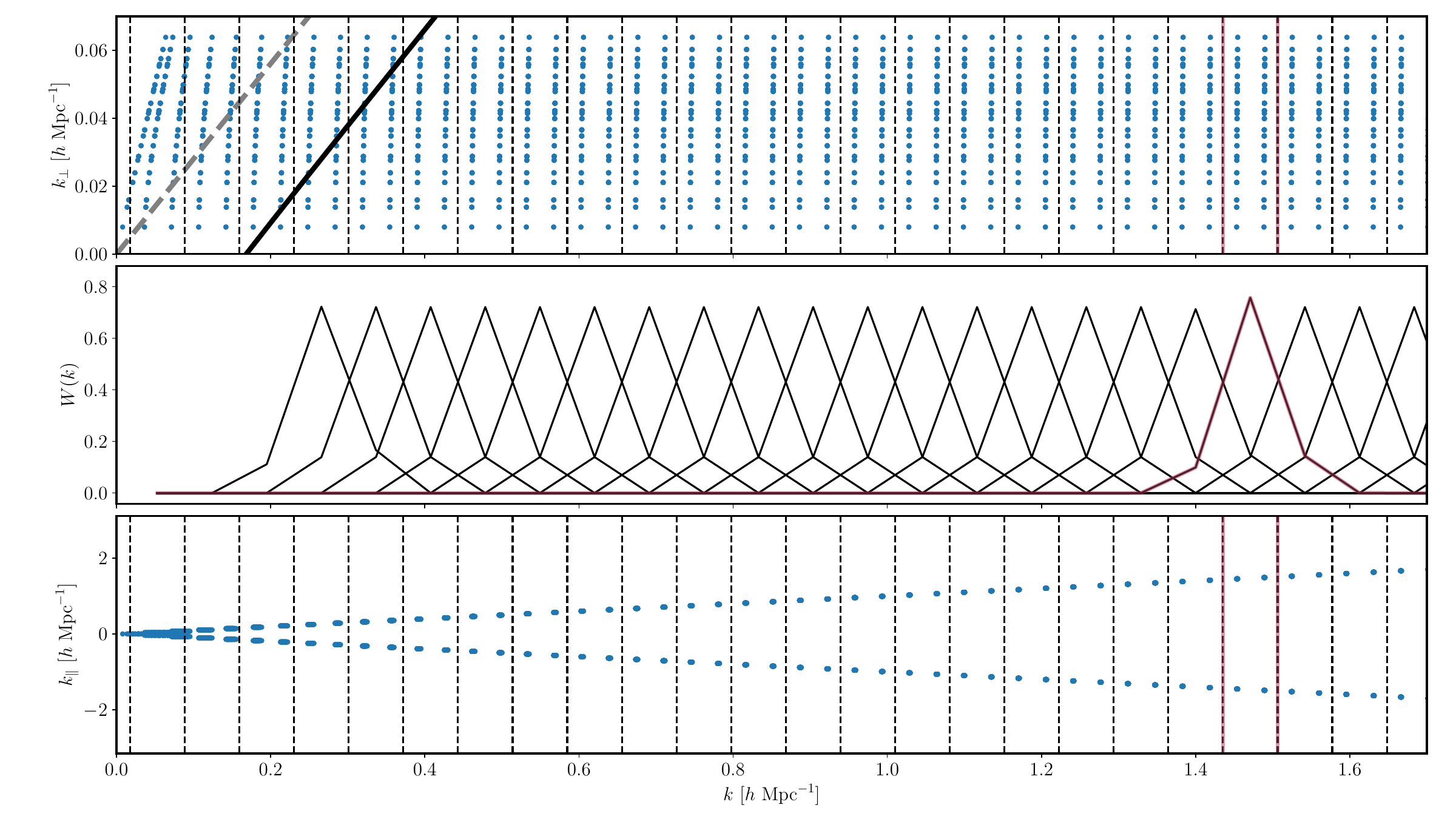}
    \caption{Distribution of baselines, and corresponding $k_\perp$ (upper panel) and delays, and corresponding $k_\parallel$ (lower panel) per spherical bin and window function for Band~2 \citep{HERA_new_upper_limits}. Vertical dashed lines represent bin edges. The dashed tilted line represents the horizon, whilst the solid tilted line represent the cut taken in the analysis (the wedge buffer, see Sec.~\ref{sec:data}). Note that several baselines can be stacked on a single dot. The outlier, centred on $k=1.47~h\mathrm{Mpc}^{-1}$, which is taller than the other window functions, is highlighted in red. It is investigated in Appendix~\ref{app:outlier}.}
    \label{fig:baselines_in_bins}
\end{figure*}

\subsection{Cylindrical and spherical window functions} \label{subsec:3_delay}

We consider the full Phase I HERA data set introduced in Sec.~\ref{subsec:data_hera} and compute the corresponding exact cylindrical window functions $W(b,\tau; k_\perp, k_\parallel)$ with the formalism described in Sec.~\ref{sec:methods}, for Band 1 and 2, as well as for the five fields considered in the analysis.
According to equation~\eqref{eq:def_wf}, each cylindrical window function corresponds to a baseline-delay pair $(b, \tau)$.
We show in Fig.~\ref{fig:cylindrical_wf_Band1} the result for Band 1 and a set of HERA baseline lengths and delays. We see that, although each window function reaches its maximum at the expected $(k_\perp, k_\parallel)$ given by equation~\eqref{eq:wf_centres} for each $(b, \tau)$ pair, they have a non-zero width, meaning power from neighbouring cylindrical $k$-modes will leak into the measurement of the power spectrum at a given $(k_\perp, k_\parallel)$. In particular, longer baselines lead to a longer tail towards low $k_\parallel$, as already observed in \citet{LiuShaw_2020}. Note that the effect will be clearer in the next section, when baselines longer than the maximum baseline in the current HERA array ($b > 120\,\mathrm{m}$) are included. 

\begin{figure}
    \centering
    \includegraphics[width=.9\columnwidth]{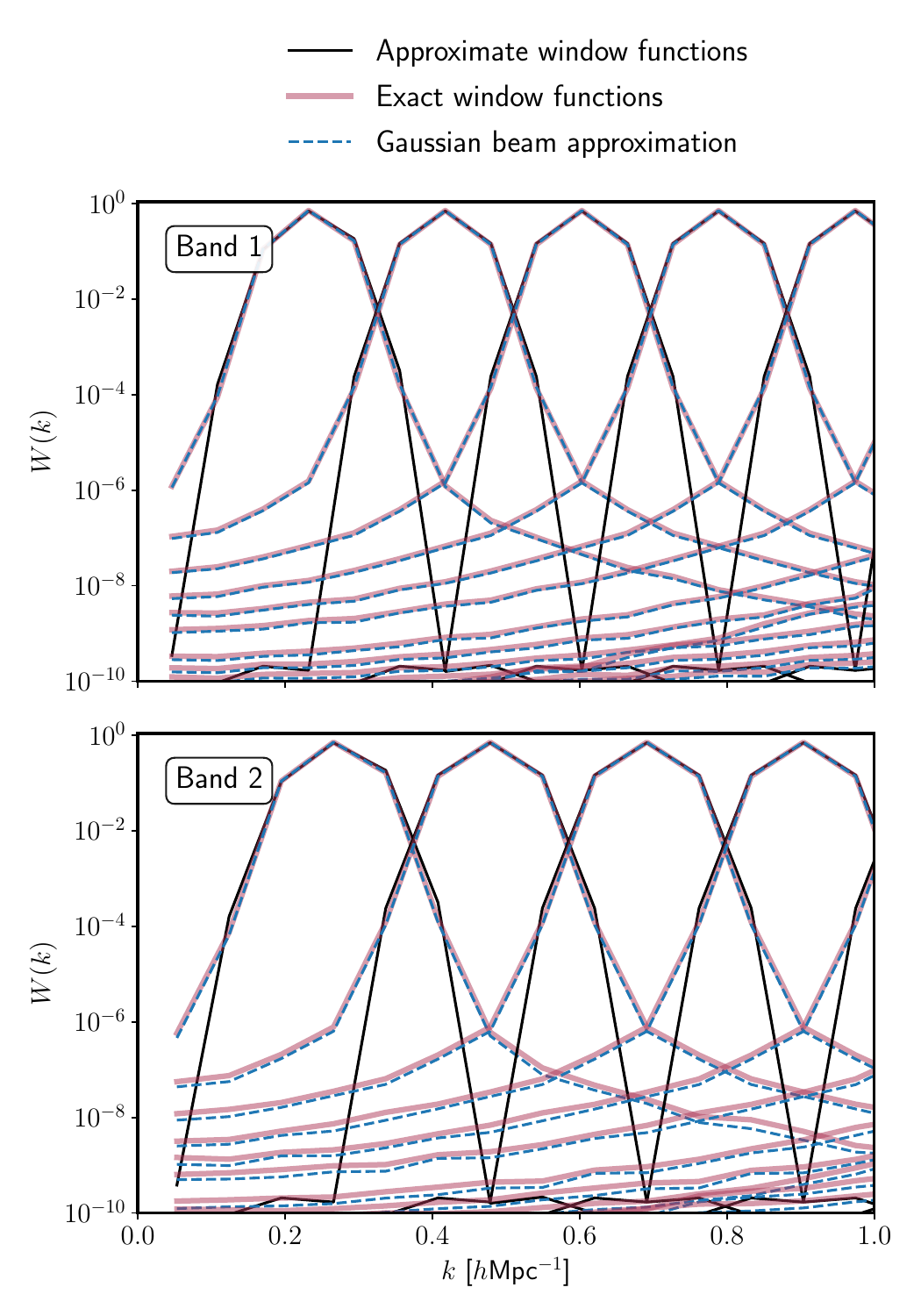}
    \caption{Exact spherical window functions obtained for the full Phase I HERA analysis of Band 1 (upper panel) and Band 2 (lower panel) \citep{HERA_new_upper_limits}, compared to the initial computation and to a case where the HERA beam is approximated by a Gaussian (Appendix~\ref{app:Gaussian_beam}). The inverse-variance weighting leads to zero window functions at the largest scales. Most of the cylindrical structure observed in Fig.~\ref{fig:cylindrical_wf_Band1} is washed out by the spherical average due to the structure of the HERA array (see text), resulting in nearly identical approximate and exact window functions. The difference induced by the frequency-dependence and the spatial structure of the beam is visible in the tails of the exact window functions, but only has a small contribution ($W(k) \lesssim 10^{-4}$).}
    \label{fig:spherical_wf}
\end{figure}

\begin{figure}
    \centering
    \includegraphics[width=.8\columnwidth]{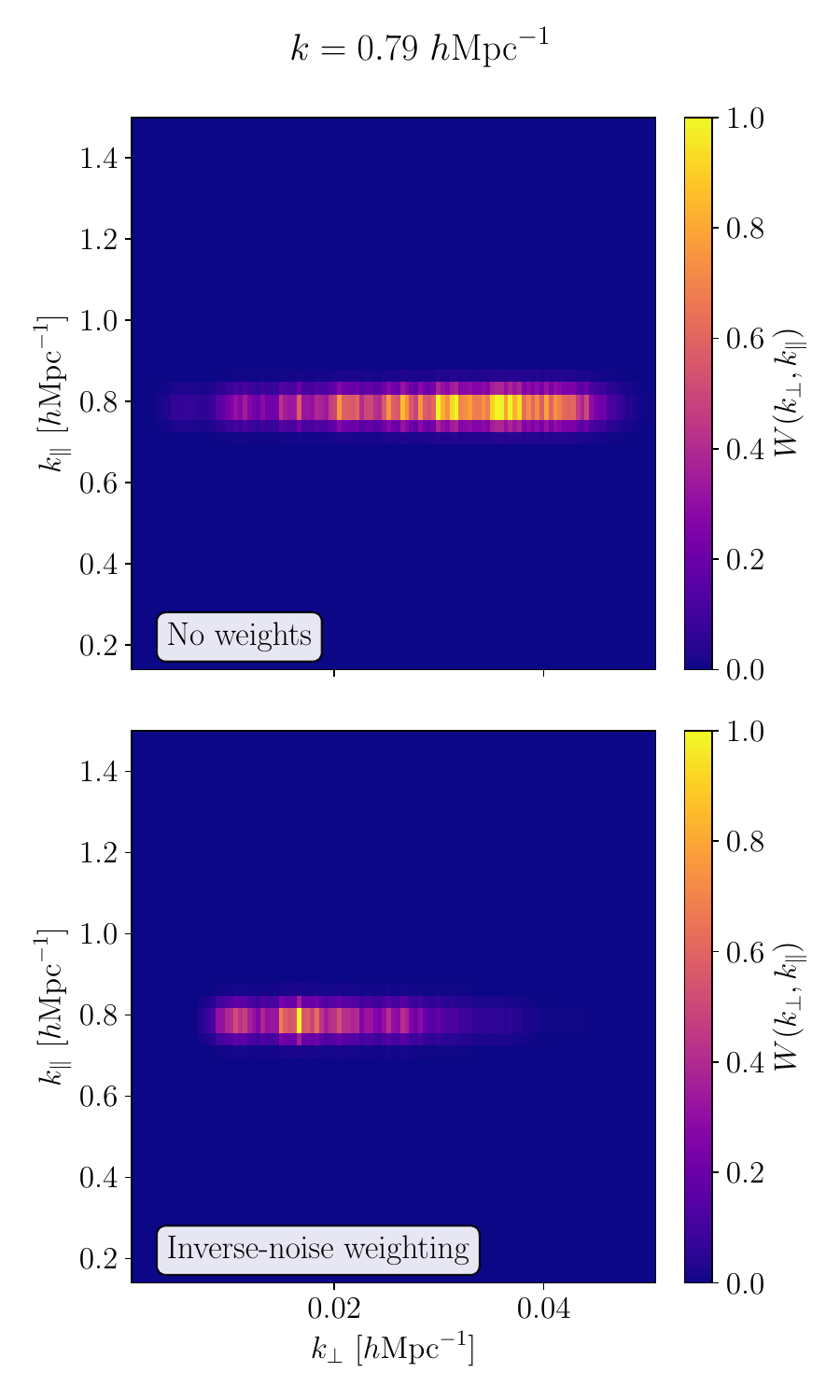}\\   
    \includegraphics[width=.8\columnwidth]{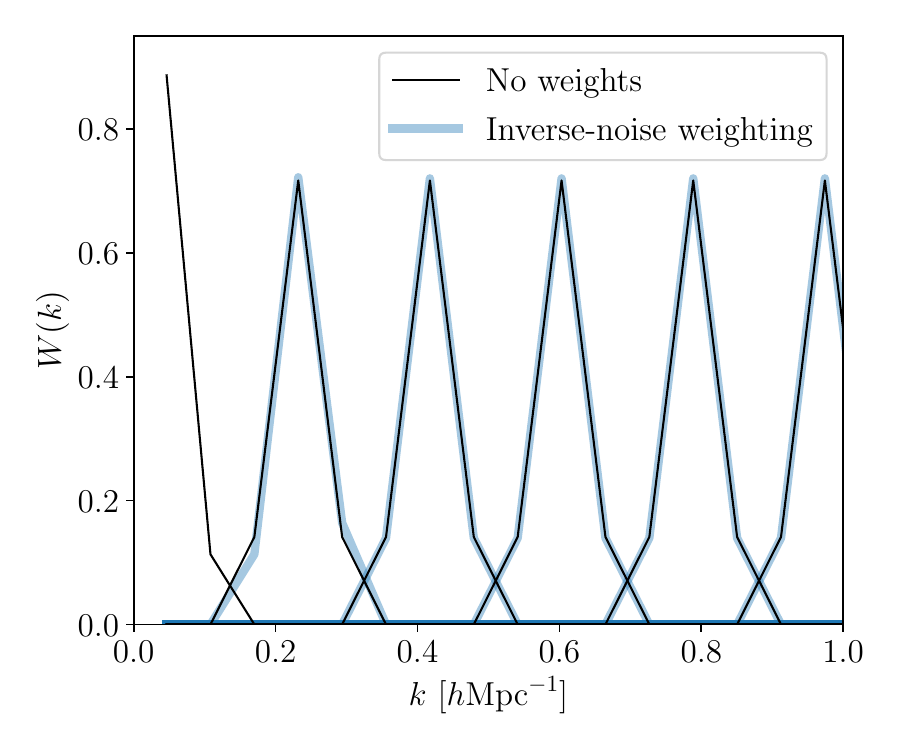}
    \caption{Perpendicular and parallel modes contributing to the spherical bin centred on $k=0.79\,h\mathrm{Mpc}^{-1}$, after summing contributions from all $(b, \tau)$ pairs. \textit{Upper panel:} No weights applied to the data. \textit{Middle panel:} After applying inverse-variance weights and removing flagged data. \{As already seen in Fig~\ref{fig:hist_weights}, most of the power is redirected toward short baselines (small $k_\perp$). \textit{Lower panel:} Resulting spherical window functions for both cases. The spherical average being almost horizontal ($k\sim k_\parallel)$, most of the cylindrical structure is lost in the average and the spherical window functions are nearly identical. Only the wedge filtering has a noticeable effect.}
    \label{fig:contribution2kbin_weights}
\end{figure}

 The spherical window functions are obtained by averaging the contribution of each $W(b,\tau)$ window to the appropriate spherical $k$-bin, as illustrated in Fig.~\ref{fig:baselines_in_bins}. The resulting spherical window functions, which were also presented in \citet{HERA_new_upper_limits}, are shown in Fig.~\ref{fig:spherical_wf}. When taking the spherical average of the four-dimensional exact window functions, we apply inverse-noise weighting and discard flagged data, as described in Sec.~\ref{subsec:data_hera} (see also Sec.~\ref{subsec:3_impact_props}), effectively cutting off the low-$k$ modes located in the foreground wedge and neighbouring buffer ($k \lesssim 0.15\,h\mathrm{Mpc}^{-1}$), as illustrated in the lower panel of Fig.~\ref{fig:contribution2kbin_weights}. The small difference observed between approximate and exact window functions can be explained by different factors which we will investigate in the next section. Notably, the baselines considered cover a range of $k_\perp$ much smaller than $k_\parallel$ ($ 5.8 \times 10^{-3} < k_\perp / [h \mathrm{Mpc}^{-1}]< 4.7 \times 10^{-2}$ for Band~1, $ 8.0 \times 10^{-3}  < k_\perp / [h \mathrm{Mpc}^{-1}] < 6.4\times 10^{-2}$ for Band~2), such that a spherical bin is roughly equivalent to a $k_\parallel$ and only one baseline is sufficient to sample the spherical $k$-space properly. Only for small $k_\parallel $ could the difference be more significant, but these modes live inside the wedge. This is illustrated in the upper panel of Fig.~\ref{fig:contribution2kbin_weights}, where we show the contribution of perpendicular and parallel modes to a single spherical $k$-bin.

This close to one-to-one mapping between a given spherical $k$ and a given $k_\parallel$ or $\tau$ can also explain the outlier window function highlighted
in Fig.~\ref{fig:baselines_in_bins}. This window function, centred on $k = 1.47\,h\mathrm{Mpc}^{-1}$, is narrower -- hence, taller, than its neighbours. Indeed, because of the normalisation in equation~\eqref{ref:eq_wf_norm}, a narrower window function will be taller, and conversely. The shape of this outlier is a symmetry effect of the spectral window cut. Indeed,  $k = 1.47\,h\mathrm{Mpc}^{-1}$ corresponds to a delay located one quarter of the way along the delay range defined by that spectral window, and the low-$k$ tail of the window function centred on $k$ will receive a zero contribution from delays located at the edge of the beam, effectively lowering the amplitude of this tail; whilst the larger modes will not (see Appendix~\ref{app:outlier} for details).

Having access to the exact cylindrical window functions is essential to an accurate theoretical interpretation of the power spectrum estimates. First, convolving theoretical models with the exact window functions will move them to the same space as the data, hence facilitating their comparison. Second, the window functions give access to the distribution of power in the data between line-of-sight and perpendicular modes, which will be useful when testing non-Gaussian models for the cosmological signal. The possibility to compare theoretical models to data at the cylindrical power spectrum level will be included in future versions of the HERA likelihood\footnote{Available at \url{https://github.com/HERA-Team/pspec_likelihood} -- under development.}.
Until now, the model testing was done at the spherical power spectrum level, where a lot of the information has been smoothed out: As seen in Fig.~\ref{fig:spherical_wf}, the approximate and exact spherical window function only differ for modes far from the centre of the window function, whose contribution to the bin is lower than $10^{-4}$. Although small, this difference can become significant in the presence of extremely bright foregrounds, e.g., near the wedge.\\

In Fig.~\ref{fig:spherical_wf}, we additionally compare our results to the exact window functions obtained when approximating the HERA beam by a Gaussian beam (see Appendix~\ref{app:Gaussian_beam}). We see that the approximation works very well in spherical space, with the tails of the window functions being underestimated by only about $5\%$. This is expected as these tails do not come from the beam's side lobes, which will be poorly reproduced by a Gaussian, but from the shape of the Fourier transform of the taper used along the spectral window. We will discuss this idea further in Sec.~\ref{subsec:3_validation}. In cylindrical space, because the approximated Gaussian beam is slightly wider but decreases much more steeply (exponentially) than the HERA beam (Fig.~\ref{fig:fit_Gaussian_beam}), the Gaussian window functions are wider than the exact ones by about $\Delta k_\perp = 0.01~h\mathrm{Mpc}^{-1}$, a difference that is washed out by the spherical average. Along $k_\parallel$, because of the Gaussian beam being steeper, the fluctuations corresponding to the Fourier transform of the tapering function are amplified. However, this effect occurs only for contributions $W(k)<10^{-6}$. 

Being able to approximate the beam by a Gaussian is extremely useful. First, because all calculations outlined in Sec.~\ref{subsec:2_wf} can be done analytically, avoiding numerical issues such as resolution or sampling limits, as well as significantly lowering computing times. Second, because accurately characterising the beam of an instrument is an extremely difficult exercise. Different approaches have been used until now, including simulations \citep{TrottdeLeraAcedo_2017, FagnonideLeraAcedo_2021_PhaseI, FagnonideLeraAcedo_2021_Vivaldi}, and in-situ measurements \citep{PupilloNaldi_2015, NebenBradley_2015, JacobsBurba_2017, LineMcKinley_2018, NunhokeeParsons_2020}, often with limited precision, especially on the structure of the side lobes.

\subsection{Impact of different elements on the window functions}
\label{subsec:3_impact_props}

In this section, we investigate how the instrument characteristics, as well as analysis choices, can impact the window functions and, in turn, the estimated power spectrum. To ease computations, we consider a Gaussian beam instead of the simulated HERA beam (see Appendix~\ref{app:Gaussian_beam}).

\subsubsection{Weights}

We show in the top panel of Fig.~\ref{fig:contribution2kbin_weights} the cylindrical window functions obtained after adding all the $(b, \tau)$ pairs contributing to the spherical bin centred on $k=0.79\,h\mathrm{Mpc}^{-1}$. In the middle panel, we show the result of applying inverse-variance weights and removing flagged data on the composition of the bin.
As in Fig.~\ref{fig:hist_weights}, we see that applying weights lowers the contribution of long baselines (sampling larger $k_\perp$) and enhances the contribution of short baselines. Indeed, because of the high redundancy level of the HERA array, the power probed by short baseline lengths is sampled by many more antenna pairs than long baselines, leading to lower noise.

The cylindrical window functions shown in the upper and middle panels are then spherically averaged, effectively averaging along horizontal lines of constant $k_\parallel$, since $k_\perp \ll k_\parallel$, and, in turn, $k \sim k_\parallel$, and the resulting spherical window functions are shown in the lower panel. Because of this average, the difference between the non-weighted and weighted case is mostly washed out, leading to almost identical spherical window functions. Again, the difference is only seen for $W(k)\leq10^{-6}$ contributions. Only for low $k$-bins does the weighting introduce a significant difference in the spherical window functions, because of the effective removal of the contribution from short baselines.

This example demonstrates again the importance of cylindrical window functions in future, high precision theoretical interpretations of 21\,cm power spectrum measurements.

\subsubsection{Spectral properties}

\begin{figure}
    \centering
    \includegraphics[width=.9\columnwidth]{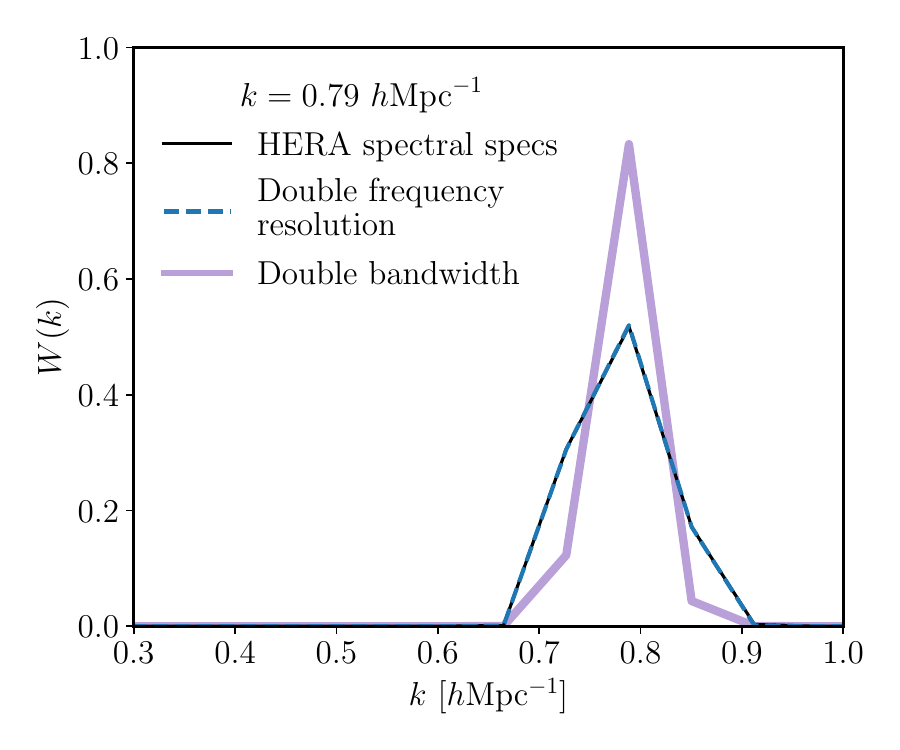}
    \caption{Spherical window function centred on $k=0.79~h\mathrm{Mpc}^{-1}$ at $z=10.4$, obtained with 50 mock baselines ranging from 10 to 230\,m long. We consider different spectral parameters and characteristics of the instrument. The addition of long baselines (not present in the HERA array) explains why the window functions presented here are wider than Fig.~\ref{fig:spherical_wf}. Increasing the frequency resolution does not have an impact on the window functions, whilst increasing the bandwidth helps narrowing them down. These two effects are explained by a simple Fourier sampling argument (see text).}
    \label{fig:spectral_dep_wf}
\end{figure}

We compute the spherical window functions with the Gaussian beam, for different choices of spectral windows, all analogous to Band~1, that is centred on $z=10.4$. We choose a set of 500 mock baselines, whose lengths range between 1 and 500\,m, leading to window functions very different from the ones shown in Fig.~\ref{fig:spherical_wf} for the HERA baselines ($b \lesssim 120\,\mathrm{m}$). 

\begin{figure*}
    \centering
    \includegraphics[width=\textwidth]{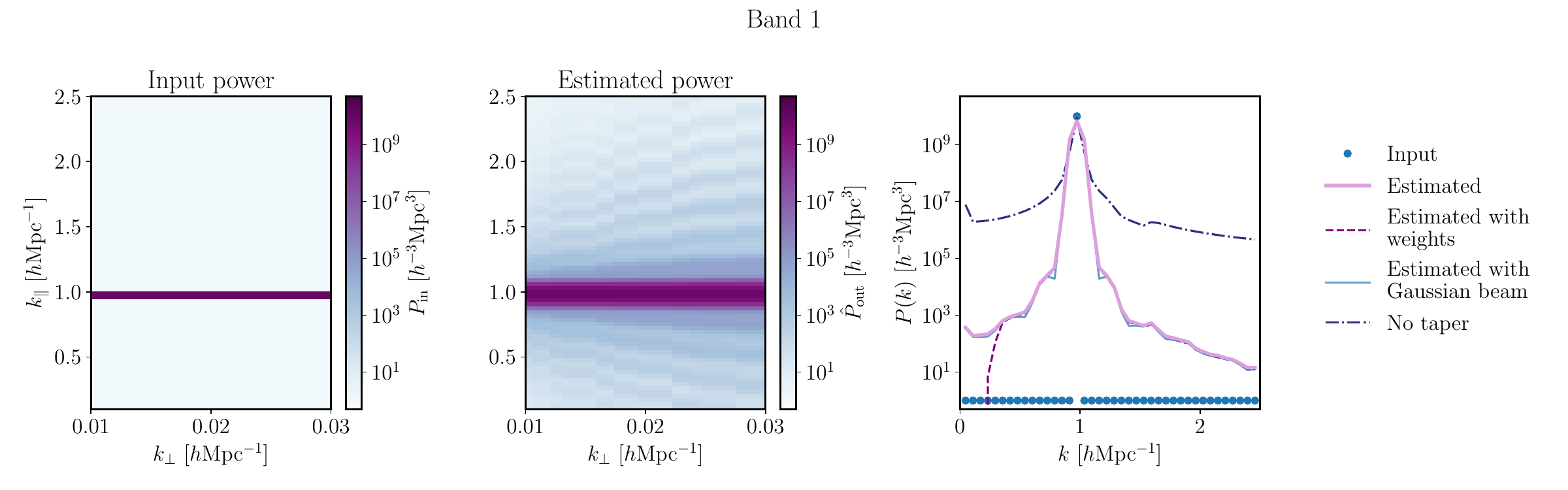} 
    \caption{Impulse response of the interferometer, illustrating mode-mixing. We compare the input (Dirac) power spectrum given in equation~\eqref{eq:dirac_input} to the resulting estimated bandpowers obtained with the HERA window functions on Band~1 by applying equation~\eqref{eq:output_ps_def}.
    \textit{Left and middle panels:} Cylindrical power spectra. \textit{Right panel:} Spherical power spectrum. In the right panel, we compare the spherical power spectra obtained in different cases: with exact window functions but with or without weights (dashed purple line, thick pink line, respectively), without a frequency taper (dash-dotted line), or with weights and taper but a Gaussian beam (thick light blue line). We see that mode mixing leads to power leaking outside of the impulse, and so missing power in the estimator at $k=k_0$. This effect is, however, mitigated by the use of a taper, which however adds structure to the recovered cylindrical power (see the ripples in the middle panel). Again, the Gaussian approximation of the beam performs well.}
    \label{fig:cylind_estimated_Dirac}
\end{figure*}

We show, in Fig.~\ref{fig:spectral_dep_wf}, the window function centred on $k=0.79~h\mathrm{Mpc}^{-1}$ for the HERA spectral specs (thin solid line). As expected, including longer baselines leads to enhanced low-$k$ tails for all bins. 
We also show the window function obtained for the same bandwidth as Band~1, but doubling the frequency resolution, that is for a channel width of $48.83\,\mathrm{kHz}$ instead of $97.66\,\mathrm{kHz}$ for HERA \citep{DeBoerParsons_2017} (dashed line). Both lines perfectly overlap, mostly because increasing the frequency resolution is only equivalent to extending the range of $k_\parallel$ covered to larger values but will not impact $k$-bins already covered.
On the other hand, increasing the bandwidth does not change the range of $k_\parallel$ probed but their sampling resolution. We double the bandwidth, now including frequencies $109.2 \leq \nu/\mathrm{MHz} < 141.2$, and show the result as the thick purple line. We show that the resulting window function is narrower than for the shorter bandwidth, which can be traced back to the cylindrical window function being narrower along $k_\parallel$. This is expected since the Fourier transform of a wider taper will be narrower. We find the cylindrical window function to be wider along $k_\perp$ for a wider bandwidth. This is also expected since the beam shift illustrated in Figs.~\ref{fig:Gbeam_shift} and \ref{fig:freq_vs_kperp} will be more significant on a longer range of frequencies. Hence, choosing a longer bandwidth can be an interesting strategy to narrow window functions and limit foreground leakage from neighbouring modes. However, such a choice comes with theoretical drawbacks as the lightcone approximation will not hold on wide spectral windows \citep{BarkanaLoeb_2004}. Here, doubling the frequency range covered is equivalent to doubling the redshift range to $\Delta z = 2.95$, corresponding to 172\,Myr. Ignoring the line-of-sight evolution of the cosmological signal is effectively equivalent to underestimating (overestimating) the 21\,cm power on large (small) scales \citep{DattaMellema_2012} and could have strong implications on the theoretical interpretation of the observations.

\subsubsection{Spherical binning}

The choice of spherical bins used to derive upper limits from the initial Phase I data set \citep{HERA_upper_limit_H1C_IDR2} has been changed for the newest results, based on the full Phase I data release \citep[see Sec.~\ref{subsec:data_hera} and][]{HERA_new_upper_limits}. These new spherical bins allow an even distribution of $(k_\parallel, k_\perp)$ pairs probed in each spherical bin, as illustrated in Fig.~\ref{fig:baselines_in_bins}. The upper and lower panels of this figure present the distribution of baseline-delay $(b, \tau)$ pairs contributing to each spherical $k$-bin for Band~2. As mentioned before, in the HERA data, $k \sim k_\parallel$, which explains the nearly vertical distribution of $(b, \tau)$ pairs in the upper panel and the nearly horizontal distribution in the lower panel. Indeed, all baselines will contribute to each $k$-bin, for a given $k_\parallel$, or, equivalently, $\vert \tau \vert$. We show the corresponding spherical window functions obtained for weighted HERA data in the middle panel.

To avoid sampling errors, it is necessary to make sure at least one delay is included in each $k_\parallel$-bin when choosing the grid the window functions will be computed along. The spacing $\Delta k_\parallel$ between two modes must be a multiple of $2 \pi \Delta \tau / \alpha(z)$ where $\Delta \tau$ is the spacing between two measured delays, that is $1/B$ where $B$ is the length of the spectral window considered.
Because, for HERA, the $k_\perp$ sampled by the instrument are shorter than the $k_\parallel$ by at least one order of magnitude, choosing the spherical binning is effectively equivalent to choosing the $k_\parallel$ binning. Hence, to not oversample the $k$-range, one must ensure that $\Delta k$ is a multiple of $2 \pi /B \alpha(z)$.\\

The above examples show what impact some analysis choices, such as the bandwidth or the $k$-sampling pattern, can have on the window functions and, subsequently, on power spectrum measurement and analysis. In the following section, we demonstrate for several test cases the importance of window functions when reconstructing the 21\,cm power spectrum via a delay-based analysis.

\subsection{Validation} \label{subsec:3_validation}

In this section, we illustrate how window functions can explain mode mixing by analysing the impulse response of the power spectrum estimator. We then proceed to applying the exact window functions to the validation simulations presented in \citet{HERA_validation_2022}, for a data set made of a known cosmological signal, following a power-law, as well as physical foregrounds (see Sec.~\ref{subsec:3_validation} for details).

\subsubsection{Toy models to illustrate mode mixing}

As a proof of concept, let us consider the impulse response of the power spectrum estimator. That is, we consider a spherical power spectrum such that
\begin{equation} \label{eq:dirac_input}
P_\mathrm{in}(k) = 
\left\{ \begin{aligned}
    & 10^{10} &\text{if}\, k=k_0, \\
    & 1 &\text{else},
    \end{aligned}
    \right.
\end{equation}
and construct the corresponding power in cylindrical space for the HERA setup, $P_\mathrm{in}(k_\perp, k_\parallel)$, shown in the left panel of Fig.~\ref{fig:cylind_estimated_Dirac}. We then use equation~\eqref{eq:def_wf} to obtain the estimated bandpowers for the input power spectrum:
\begin{equation}\label{eq:output_ps_def}
\hat{P}_\mathrm{out}(b,\tau) =  \int \mathrm{d} k_\perp \mathrm{d} k_\parallel  \,P_\mathrm{in}(k_\perp, k_\parallel)\,W(k_\perp, k_\parallel;b,\tau).
\end{equation}
The estimator at $(b, \tau)$ is then matched to the appropriate $(k_\perp, k_\parallel)$ pair according to equation~\eqref{eq:wf_centres}.
Results for Band~1 are shown for the cylindrical and spherical power spectra in, respectively, the middle and right panels of Fig.~\ref{fig:cylind_estimated_Dirac}. We compare several results in spherical space: The power spectrum recovered with window functions including or not the data noise errors and flagging (see Sec.~\ref{subsec:data_hera}), in the dashed purple and thick pink line, respectively. We also show the results obtained with weighted window functions for a Gaussian beam and for the HERA beam, but without applying a tapering function along the spectral window.

We see that despite the input power being overall well recovered, the estimated power spectrum contains power at scales $k \neq k_0$, which is a perfect illustration of mode mixing \citep{MoralesHazelton_2012}. The shape of the recovered spherical power spectrum is the exact shape of the spherical window function at the corresponding $k$-bin, scaled by the amplitude of the input power. We find that the power at $k=k_0$ is underestimated by a factor three whilst the power in neighbouring cylindrical cells is overestimated by a factor eight. As already noted in Sec.~\ref{subsec:3_impact_props}, applying inverse-variance weights to the window function only has a small impact on the recovered power, removing the contribution from low-$k$ modes, or small delays, located in the foreground wedge. Indeed, the weights change the sampling pattern along $k_\parallel$ (see Fig.~\ref{fig:contribution2kbin_weights}), to which the spherical window functions are only weakly sensitive. The Gaussian approximation once again performs well, slightly enhancing the modulations around the peak due to the Gaussian beam power decreasing exponentially with $k_\perp$, in contrast to the simulated HERA beam.

These modulations, also seen as ripples along $k_\parallel$ in the recovered cylindrical power spectrum, correspond to the Fourier transform of the beam being convolved by the Fourier transform of the tapering function, a leakage that was already observed in \citet{HERA_validation_2022}. If no taper (or a step-like taper) is applied, the Dirac input power spectrum is smeared into a $ k_\parallel^{-2}$ power-law corresponding to the Fourier transform of the step function -- the \texttt{sinc} function. This is visible in the right panel, where we show the spherical power spectrum recovered in the no-taper case. According to \citet{Tegmark_1997}, a way to mitigate this effect would be to improve the modelling of the data covariance matrix, and doing so equivalently lower the weights applied to the edges of the spectral window. Additionally, applying a taper such as Blackman-Harris widens the tails of the window functions as it reduces the effective bandwidth by half (see Sec.~\ref{subsec:3_impact_props}). We refer the interested reader to \citet{ThyagarajanUdayaShankar_2013} for a discussion of the impact of tapering choices in terms of foreground leakage.\\

\begin{figure}
    \centering
    \includegraphics[width=\columnwidth]{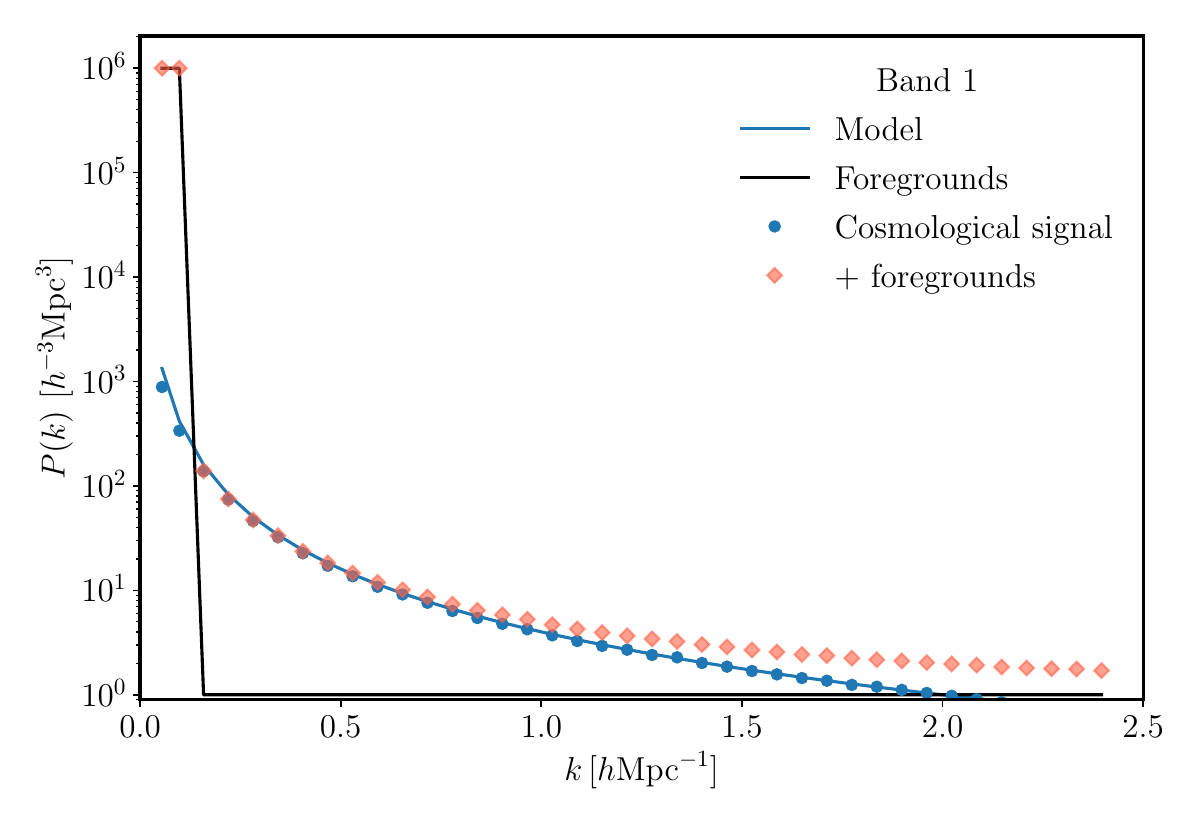}
    \includegraphics[width=\columnwidth]{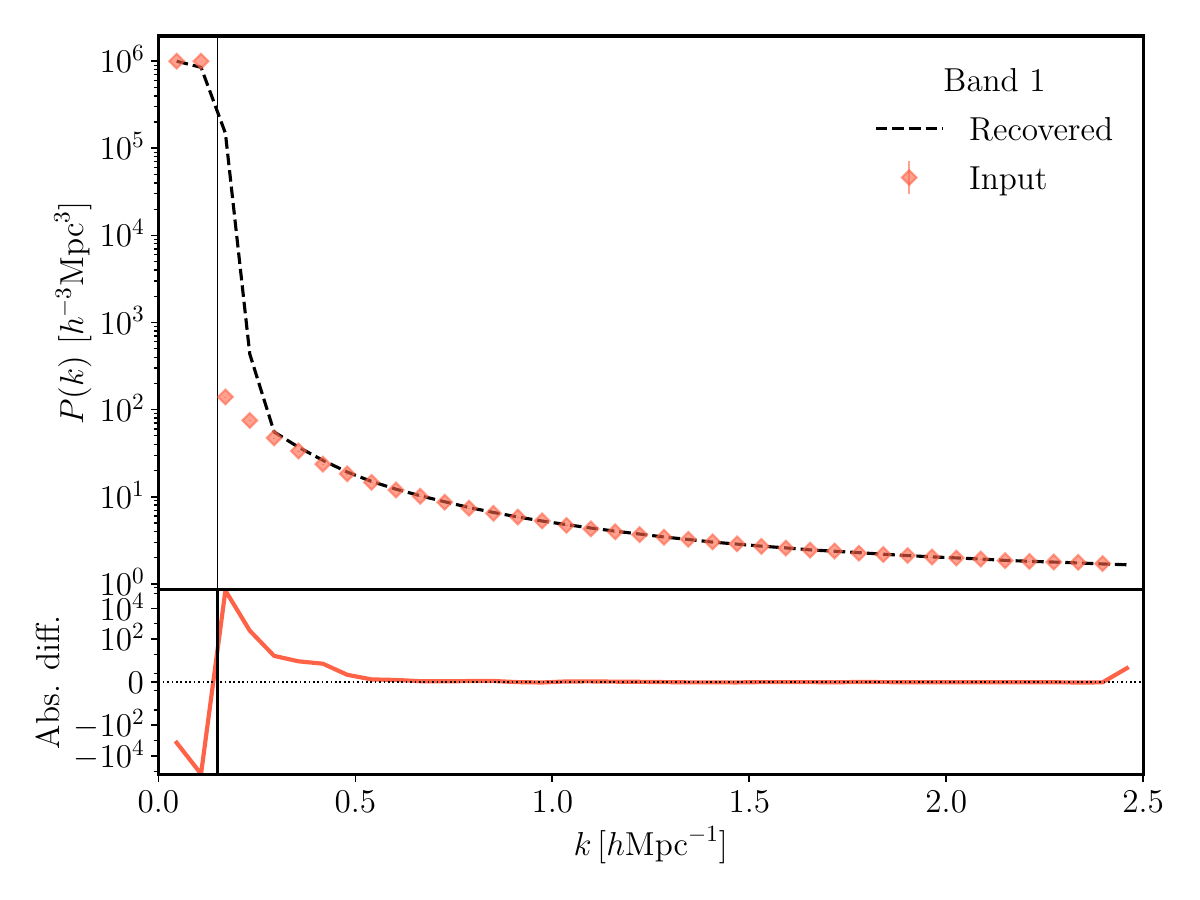}
    \caption{Input (upper panel) and recovered (lower panel, equation~\ref{eq:output_ps_def}) spherical power spectra for a toy model including a Gaussian cosmological signal and a simplified (diffuse) foregrounds model where the foregrounds are limited to $k_\parallel < k_\mathrm{lim}$, represented as the vertical line in the lower panel. This figure illustrates how the exact window functions can explain foreground leakage around the wedge.}
    \label{fig:toy_model}
\end{figure}

Let us now increase the complexity of the model and consider a simplified foregrounds and cosmological model, defined on the two HERA Bands. We generate a Gaussian cosmological signal in 3D Fourier space such that $P_\mathrm{cosmo}(k) \propto k^{-2}$, following \citet{HERA_validation_2022}. We add a simplified foreground model, analogous to a diffuse sky model, such that 
\begin{equation}
P_\mathrm{fg}(k_\perp, k_\parallel) = 
\left\{ \begin{aligned}
    & 10^6 &\text{if}\, k_\parallel<k_\mathrm{lim}, \\
    & 1 &\text{else},
    \end{aligned}
    \right.
\end{equation}
where $k_\mathrm{lim}=0.15\,h\mathrm{Mpc}^{-1}$, corresponding to the wedge limit (see \ref{subsec:data_hera}).
These two contributions, along with their sum, which is our mock signal, are shown as spherical power spectra in the upper panel of Fig.~\ref{fig:toy_model} for Band~1. Because there are no correlations between the cosmological and foreground signals, the two power spectra simply add up: $P_\mathrm{in}(k) = P_\mathrm{cosmo}(k)+P_\mathrm{fg}(k)$.

Again, we use the exact window functions and equation~\eqref{eq:output_ps_def} to obtain the estimated bandpowers corresponding to the HERA analysis. We compare in the lower panel of Fig.~\ref{fig:toy_model} the difference between the input $P_\mathrm{in}(k)$ and the output $P_\mathrm{out}(k)$ spherical power spectra. Some leakage of the foregrounds power above $k_\mathrm{lim}$ is visible in cylindrical and spherical space, extending to $\sim 2k_\mathrm{lim} = 0.30\,h\mathrm{Mpc}^{-1}$. Note that the reconstruction is identical for a Gaussian beam.

\subsubsection{Validation simulations}

\begin{figure}
    \centering
    \includegraphics[width=\columnwidth]{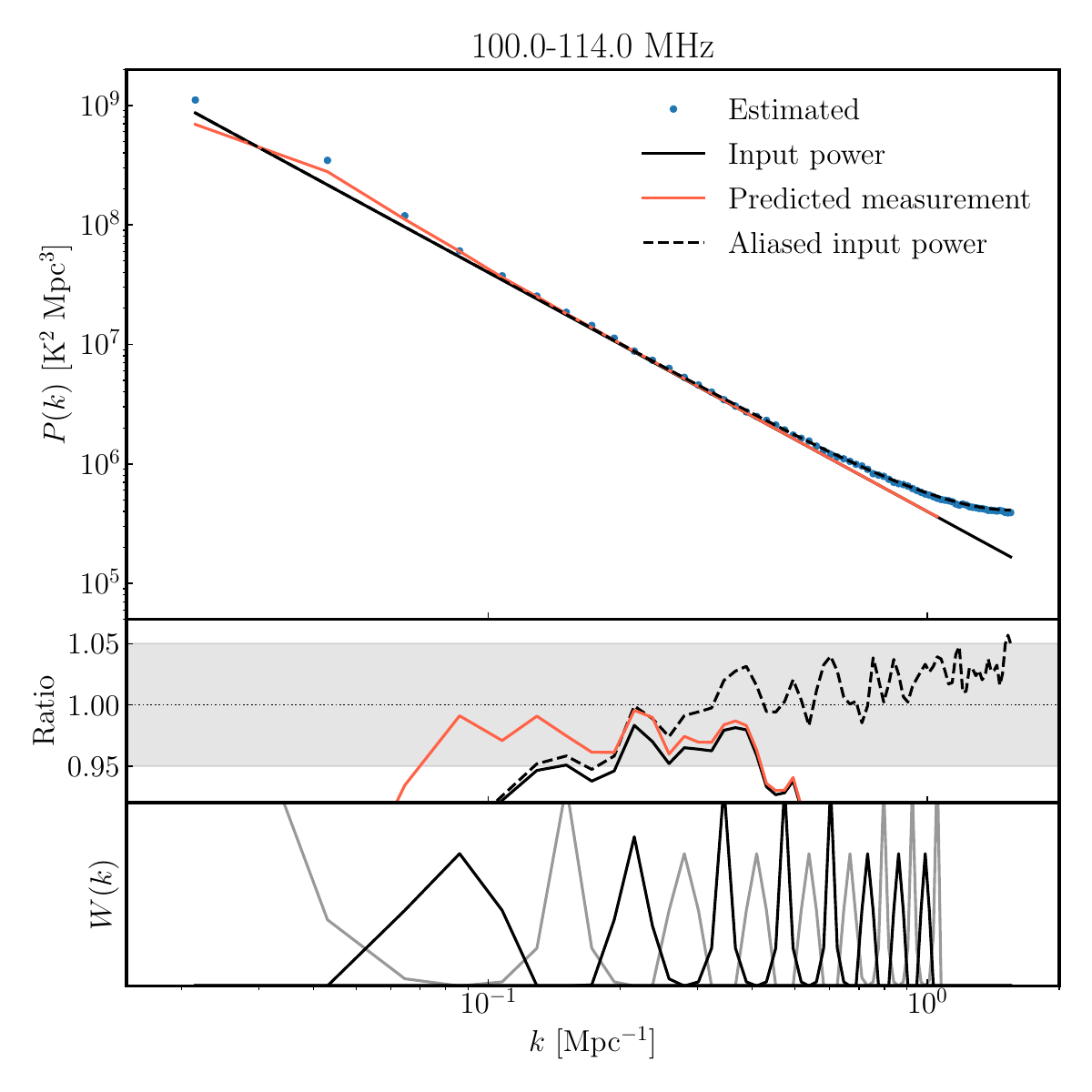}
    \caption{Comparison between the input power-law power spectrum (solid black line), and the recovered one (data points), obtained by running simulated visibilities through the \texttt{hera\_pspec} pipeline over a $100 < \nu/\mathrm{MHz} < 140$ bandwidth. Accounting for aliasing \citep[see Appendix~B of][]{HERA_validation_2022} corrects for the recovery bias on small scales ($k \gtrsim 0.4\,\mathrm{Mpc}^{-1}$, dashed line) whilst including the window function weighting (equation~\eqref{eq:def_wf}, shown on the lower panel in alternate colours) improves results on larger scales (solid red line).} 
    \label{fig:validation_result}
\end{figure}

We now consider the validation simulations introduced in \citet{HERA_validation_2022} and succinctly described in Sec.~\ref{subsec:data_sims}. We take the data set made of the mock EoR signal only, noting that it is simulated to have a power spectrum $P(k) \propto k^{-2}$. No foreground emission, noise, or instrumental corruption beyond the beam is included in the data. We consider a bandwidth $100.0 < \nu/\mathrm{MHz} < 114.0$ centred on $z=12.3$. 

These simulated visibilities are run through the HERA analysis pipeline, and Fig.~\ref{fig:validation_result} presents the resulting power spectrum (blue data points) compared to the theoretical input power (black line).
As in \citet{HERA_validation_2022}, we find that the recovered power spectrum is generally in good agreement with the theoretical $P(k) \propto k^{-2}$ on the central modes of the $k$-range. Namely, this test validates the normalisation conventions and cosmological conversions carried out throughout the analysis. However, there is a clear discrepancy on large and small scales. On small scales, the authors of \citet{HERA_validation_2022} find that the positive bias can be corrected by a simple approximation of the aliasing effect, as illustrated in Fig.~\ref{fig:validation_result} as the dashed line. This correction brings the recovery back to a $5\%$ precision on $k \gtrsim 0.4\,\mathrm{Mpc}^{-1}$. On large scales,
the discrepancy is partly due to the window functions: For each $k$-bin, the estimated power spectrum is effectively a weighted average of the true power spectrum over neighbouring modes. Since the input power spectrum is a decreasing function of $k$, this effect is stronger for low $k$-modes, corresponding to a larger power. We use equation~\eqref{eq:def_wf} to correct for this effect by including the window function weighting in the estimated power. We achieve a better recovery of the input power spectrum, reducing the discrepancy that was seen in \citet{HERA_validation_2022}: All modes $k > 0.04\,\mathrm{Mpc}^{-1}$ are recovered within a $5\%$ precision. However, the asymmetry on the window functions on the edge of the $k$-range lead to a largely underestimated power for the first few bins.
Note that, in this example, only one realisation of the mock EoR signal is used, but an even better precision can be achieved by averaging over several realisations: In \citet{HERA_validation_2022}, the aliasing corrections averaged over 50 realisations leads to a better than $1\%$ precision on the recovered power.

\section{Asymmetric window functions} \label{sec:3_asym}

\begin{figure}
    \centering
    \includegraphics[width=.9\columnwidth]{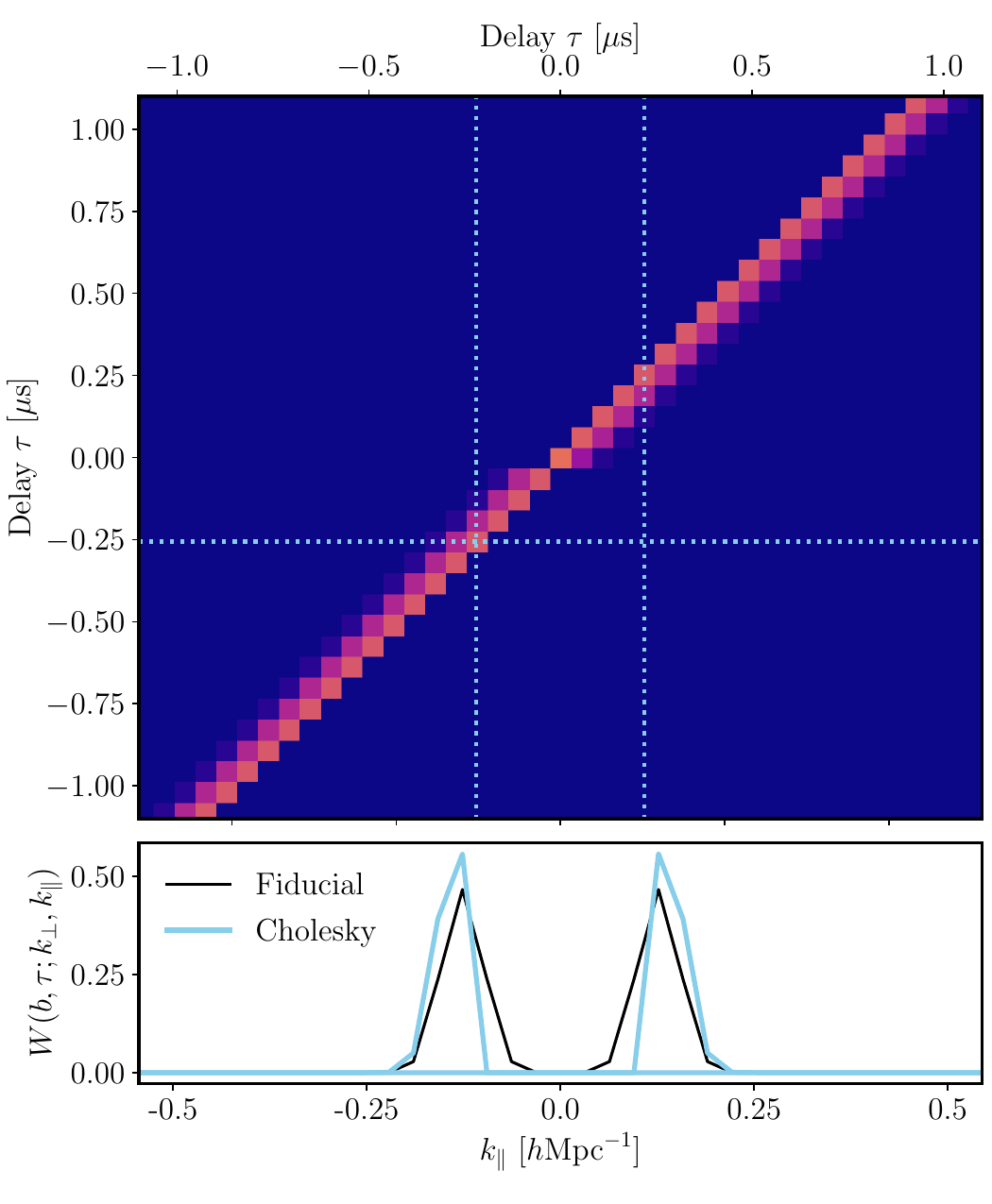}
    \caption{\textit{Top panel:} The window function matrix obtained through the Cholesky decomposition of the response matrix $\textbf{H}$ for $b=44.1\,\mathrm{m}$. \textit{Lower panel:} Rows of the window function matrix corresponding to $\vert \tau \vert = 0.26\,\mu\mathrm{s}$, for the Cholesky decomposition and the fiducial case. The symmetry of the matrix with respect to $\tau=0\,\mathrm{s}$ is a requirement to keep the window function asymmetric after folding the delay power spectra.}
    \label{fig:asym_wf_matrix}
\end{figure}

In this section, we investigate the potential of asymmetric window functions, that is window functions with a deflated low-$k$ tail, to mitigate foreground leakage around the wedge. To do so, we modify the normalisation matrix included in the analysis (see Sec.~\ref{subsec:2_estimator}) to change the shape of the window functions and obtain a window function matrix that is upper triangular. Note that this step is fully independent of the exact window functions mentioned in previous paragraphs: The exact window functions are an intrinsic effect of the data going through the instrument, whilst the analysis choices we make here are, in contrast, applied to already squared data, that is already formed delay power spectrum estimators.

We perform a Cholesky decomposition of the response matrix $\textbf{H}$\footnote{Because $\textbf{H}$ is symmetric and positive-definite, the Cholesky decomposition is unique.} introduced in Sec.~\ref{subsec:2_estimator}, that is we can write $\textbf{H} = \textbf{L} \textbf{L}^\dagger$, where $\textbf{L}$ is a lower triangular matrix with real and positive diagonal entries, and the dagger denotes the conjugate transpose of $\textbf{L}$. Identifying with the terms of equation~\eqref{eq:WMH} and observing that all the terms in $\textbf{L}$ are real, we have $\textbf{M} = \textbf{L}^{-1}$ and $\textbf{W}=\textbf{L}^\mathrm{t}$, the transpose of $\textbf{L}$. We adjust the normalisation of each row of $\textbf{M}$ to ensure that the resulting window functions sum to one for each bin. The resulting asymmetric window function matrix in instrument space is presented in Fig.~\ref{fig:asym_wf_matrix}.

There are some technical subtleties one needs to be aware of when substituting for the new normalisation matrix in the analysis. First, the two axes of $\textbf{W}$ correspond to, respectively, delay- and cosmological space, which are identified in the framework of the delay approximation. When the delay bandpowers will be binned by $k \sim \vert \tau \vert$ to form a spherical power spectrum, the window functions will effectively be folded along the delay axis. Hence, to obtain an asymmetric window function in spherical space with a smaller low-$k$ tail, one must define a block window function matrix made of two blocks: A lower triangular block for negative delays, and an upper triangular for positive delays. This structure is clearly visible in Fig.~\ref{fig:asym_wf_matrix}. Second, the normalisation of $\textbf{M}$ must be adjusted to ensure the normalisation of the window functions as in equation~\eqref{eq:norm_wf}. 

\begin{figure}
    \centering
    \includegraphics[width=\columnwidth]{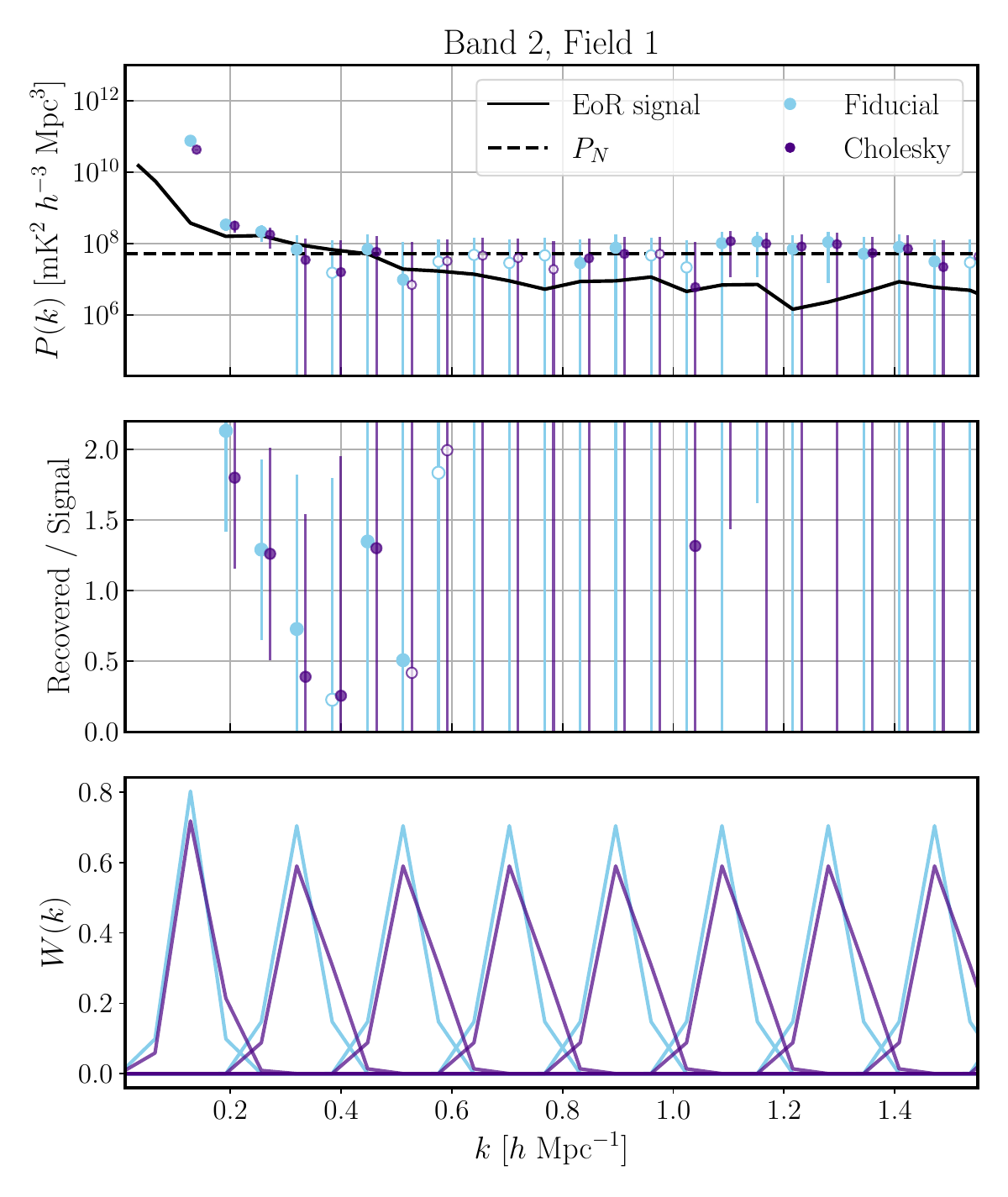}
    \caption{\textit{Upper panel:} Recovered power spectra after running the full validation simulations of \citep{HERA_validation_2022} through the HERA analysis pipeline, including systematics treatment, for the fiducial analysis (in blue), or the Cholesky-derived asymmetric window functions (in dark purple). \textit{Middle panel:} Ratio of the recovered to the EoR power spectrum. \textit{Lower panel:} Corresponding spherical window functions. The spherical binning leads to non-zero low-$k$ tails of the asymmetric window functions, compared to Fig.~\ref{fig:asym_wf_matrix}. In all panels, open-filled plot symbols correspond to negative values of the estimated power. Because of their enhanced high-$k$ tail, the asymmetric window functions integrate down to a $k$-value slightly larger than the fiducial ones -- explaining the shift between the light and the dark symbols. In both cases, the EoR signal and the noise are recovered within error bars in the region of $k$-space where they dominate, respectively. Applying asymmetric window functions does not seem to have a significant impact on the recovered power.}
    \label{fig:asym_wf}
\end{figure}

In Fig.~\ref{fig:asym_wf}, we present the result of applying these asymmetric window functions to the validation simulations introduced in Sec.~\ref{subsec:data_sims} for Field~1 (first LST cut) and Band~2. Here, the simulations include both the foregrounds and the EoR signal, as well as instrument systematics such as thermal noise, cable reflections, antenna gains and cross-coupling. The visibilities are run through the analysis pipeline, including redundant and absolute calibration, RFI flagging, systematics removal and coherent time average. The resulting power spectra are then averaged by redundancy and spherically to obtain the estimated power spectra presented in the upper panel of the figure, for the original and the asymmetric window functions (shown in the lower panel), with $2\sigma$ error bars. These results are compared to the noise floor, obtained with equation~\eqref{eq:noise_ps}, and to the power spectrum obtained when following the same steps, but for visibilities including the EoR signal only -- and no systematics. The middle panel shows the ratio of recovered power spectrum for the full data to the EoR-only power spectrum.

As already observed in \citet{HERA_validation_2022}, in both cases, the EoR signal is recovered within error bars in the region of $k$-space where it dominates both the foregrounds and the noise ($0.17 \lesssim k / [h\mathrm{Mpc}^{-1}] \lesssim 0.33$). On smaller scales, the recovered spectra are consistent with noise, proving that the systematics have been efficiently removed and that the analysis has not produced additional biases. Note that the results presented here only correspond to a sub-set of the full validation simulations, so that the recovery statistics are not as good as the results presented in \citet{HERA_validation_2022}. 
The impact of the asymmetric window functions is difficult to assess. In the region surrounding the foreground wedge ($k \sim 0.15\,h\mathrm{Mpc}^{-1}$), the power spectrum recovered with asymmetric window functions is closer to the EoR signal: It is $50\%$ to $80\%$ smaller in amplitude than with the original window functions.\\

In the previous sections, we have demonstrated the importance of knowing the exact window functions of a given power spectrum estimator to correctly assess the amount of foreground signal leaking into the cosmological signal beyond the wedge and, in turn, make an accurate theoretical interpretation of the observations. We have attempted to mitigate this intrinsic effect by post-processing the already formed delay power spectra with asymmetric window functions, whose deflated low-$k$ tails prevent some of the foreground leakage near the edge of the wedge. However, more aggressive foreground mitigation requires a more upstream approach, including explicit filtering.

\section{Conclusions} \label{sec:conclusions}

When constructing a power spectrum estimator from low-frequency interferometric data, a proper knowledge of the mapping between instrumental and cosmological space, that is of the window functions of the power spectrum estimator, is crucial to the correct theoretical interpretation of observations. In this paper, we introduced a formalism to derive these window functions, which can be applied to any delay-based analysis (Sec.~\ref{subsec:2_wf}). We demonstrated the impact of different analysis choices on the window functions, arguing in favour of a choice of spherical bins consistent with the spectral window considered (Fig.~\ref{fig:baselines_in_bins}). Namely, we showed that including long baselines in the analysis tends to enhance the low-$k$ tails of the spherical window functions, facilitating foreground leakage outside of the wedge. On the other hand, considering wide spectral window can help narrow down the window functions and concentrate the measured power around the centre of the bin (Fig.~\ref{fig:spectral_dep_wf}). However, such a choice can bias power spectrum estimates as it is in tension with the lightcone approximation, in which the fluctuations of the cosmological power along the bandwidth are ignored \citep{DattaMellema_2012}. In a similar way, the choice of the taper used to avoid edge effects along the bandwidth has a strong impact on the window functions  \citep[e.g. Fig.~\ref{fig:cylind_estimated_Dirac},][]{ThyagarajanUdayaShankar_2013}.

We focused on the Hydrogen Epoch of Reionization Array (HERA) as a case study. We derived the window functions used in the analysis of the full Phase I data, which led to the deepest upper limits on the power spectrum of the high-redshift fluctuations of the 21\,cm signal \citep{HERA_new_upper_limits}. These window functions explain part of the discrepancy observed between a theoretical model ($P(k)\propto k^{-2}$), and the power spectrum estimated after running a realisation of this model through the HERA analysis pipeline \citep[Fig.~\ref{fig:validation_result} and][]{HERA_validation_2022}. Additionally, we showed how the exact window functions are shaped by the characteristics of the array.
HERA is designed to maximise redundancy, with many short baselines and few long ones, such that $k \sim k_\parallel$. This strategy has the advantage of limiting foreground leakage by limiting the intrinsic asymmetry of the window functions. However, most of the instrument- and data-specific structure present in the cylindrical window functions, conveyed through weights and data flagging, is lost in spherical space (Fig.~\ref{fig:contribution2kbin_weights}), illustrating the importance of confronting theoretical models with observations before performing the spherical average. Such an approach, made possible by the formalism introduced in this work, will be applied to future analyses of the HERA data. 
Finally, we find that, thanks to the structure of the array, a precise knowledge of the structure of the beam is not necessary to obtain accurate window functions, even in cylindrical space: Throughout this work, we have compared the results obtained using a beam simulation \citep[][shown in Fig.~\ref{fig:HERA_beam}]{FagnonideLeraAcedo_2021_PhaseI}\footnote{The results presented in this paper were obtained with simulations of the PAPER dipole feeds \citep{FagnonideLeraAcedo_2021_PhaseI}. They will need to be updated for the next seasons of observations done with new Vivaldi feeds \citep{FagnonideLeraAcedo_2021_Vivaldi}.} or a Gaussian approximation of the beam (Appendix~\ref{app:Gaussian_beam}), and found little to no difference (e.g. Fig.~\ref{fig:spherical_wf}).
On the other hand, the chromaticity of the beam (Fig.~\ref{fig:Gbeam_shift}) is a crucial element of the window functions. In the context of the HERA analysis, we have used simple test cases to illustrate how the frequency-dependence of the beam leads to mode mixing and foregrounds leaking from their wedge into the EoR window (Figs.~\ref{fig:cylind_estimated_Dirac} and \ref{fig:toy_model}). In order to correct for this leakage \textit{a posteriori}, we modified the power spectrum estimator to form asymmetric window functions, with deflated low-$k$ tails (Fig.~\ref{fig:asym_wf_matrix}). We find that applying this technique to simulated visibilities \citep{HERA_validation_2022, HERA_upper_limit_H1C_IDR2} can prevent some foreground leakage near the edge of the wedge, but that aggressive foreground mitigation requires upstream analysis techniques and filtering (Fig.~\ref{fig:asym_wf}).

The results presented in this paper are a step towards a better understanding of the systematics currently preventing a detection of the 21\,cm signal from the Cosmic Dawn and the Epoch of Reionization. Cylindrical window functions will be instrumental in using upper limits -- and a future detection, to constrain theoretical models of the high-redshift Universe.

\section*{Acknowledgements}

The authors thank Jordan Mirocha, Ronniy Joseph, and Ian Hothi for their valuable insight on this project, as well as for their support throughout.

This material is based upon work supported by the National Science Foundation under Grant Nos. 1636646 and 1836019 and institutional support from the HERA collaboration partners.  This research is funded in part by the Gordon and Betty Moore Foundation through Grant GBMF5212 to the Massachusetts Institute of Technology.
HERA is hosted by the South African Radio Astronomy Observatory, which is a facility of the National Research Foundation, an agency of the Department of Science and Innovation.

AG's work is supported by the McGill Astrophysics Fellowship funded by the Trottier Chair in Astrophysics as well as the Canada 150 Programme. AG's and AL's work is funded by the Canadian Institute for Advanced Research (CIFAR) Azrieli Global Scholars program. Additionally, AL acknowledges support from the New Frontiers in Research Fund Exploration grant program, a Natural Sciences and Engineering Research Council of Canada (NSERC) Discovery Grant and a Discovery Launch Supplement, a Fonds de recherche Nature et technologies Quebec New Academics grant, the Sloan Research Fellowship, and the William Dawson Scholarship at McGill. This result is part of a project that has received funding from the European Research Council (ERC) under the European Union's Horizon 2020 research and innovation programme (Grant agreement No. 948764; PB). PB also acknowledges support from STFC Grant ST/T000341/1.

We thank the anonymous referee and the journal editor for their comments, which have helped improve the quality of this manuscript.

\section*{Data Availability}

All the beam simulations and data analysis programmes used in this work are freely available available online\footnote{See \url{https://github.com/HERA-Team}.}. The window function estimation programme has been released as part of the \texttt{hera\_pspec} analysis suite on the same online repository. The full HERA Phase I data used in this work and in \citet{HERA_new_upper_limits} is available upon request to the collaboration.


\bibliographystyle{mnras}

\bibliography{biblio} 



\appendix

\section{Approximating the HERA beam as a Gaussian}
\label{app:Gaussian_beam}

\begin{figure}
    \centering
    \includegraphics[width=.8\columnwidth]{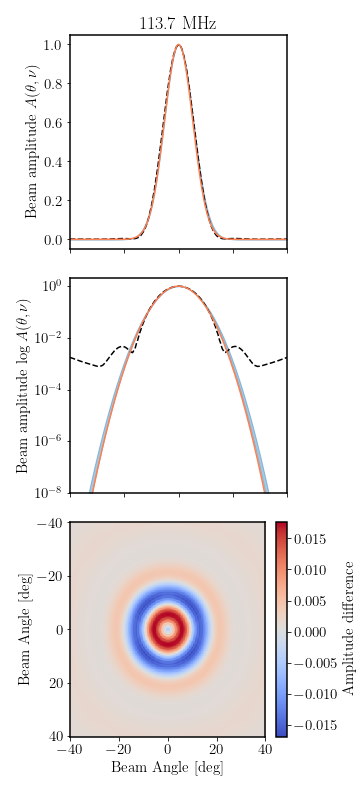}
    \caption{Result of fitting a Gaussian to the HERA beam (solid lines) for pseudo-Stokes I polarisation at $113.7\,\mathrm{MHz}$. The shaded areas represent $68\%$ confidence levels. The bottom panel shows the absolute difference between the two beams. Additionally shown in the two upper panels in orange is the beam obtained with the equation~\eqref{eq:beam_fit}.}
    \label{fig:fit_Gaussian_beam}
\end{figure}

To facilitate computations and to not be limited by the spatial resolution of the beam simulation \citep{FagnonideLeraAcedo_2021_PhaseI}, we model the HERA beam by a Gaussian. For each frequency along the HERA bandwidth ($100 \leq \nu / \mathrm{MHz} \leq 200$) and each polarisation channel, we fit a two-dimensional Gaussian to the simulated beam. The results for pI polarisation at $\nu =117.7\,\mathrm{MHz} $ are given in figure~\ref{fig:fit_Gaussian_beam} for pseudo-Stokes I polarisation\footnote{The pseudo-Stokes visibilities are a linear sum of the linear polarisation channels \citep{HamakerBregman_1996}. They can be thought of as approximations to the true Stokes visibility one would form by Fourier transforming the true Stokes parameter from the image plane to the $uv$ plane \citep{KernDillon_2020}.}. If the model seems satisfying in linear scale, the logarithmic scale shows that the amplitude of the side lobes of the beam are largely under-estimated by the Gaussian, which will have a significant impact on measurements of large amplitude signals such as foregrounds. However, we do not expect the impact to be large on window functions.

\begin{figure}
    \centering
    \includegraphics[width=\columnwidth]{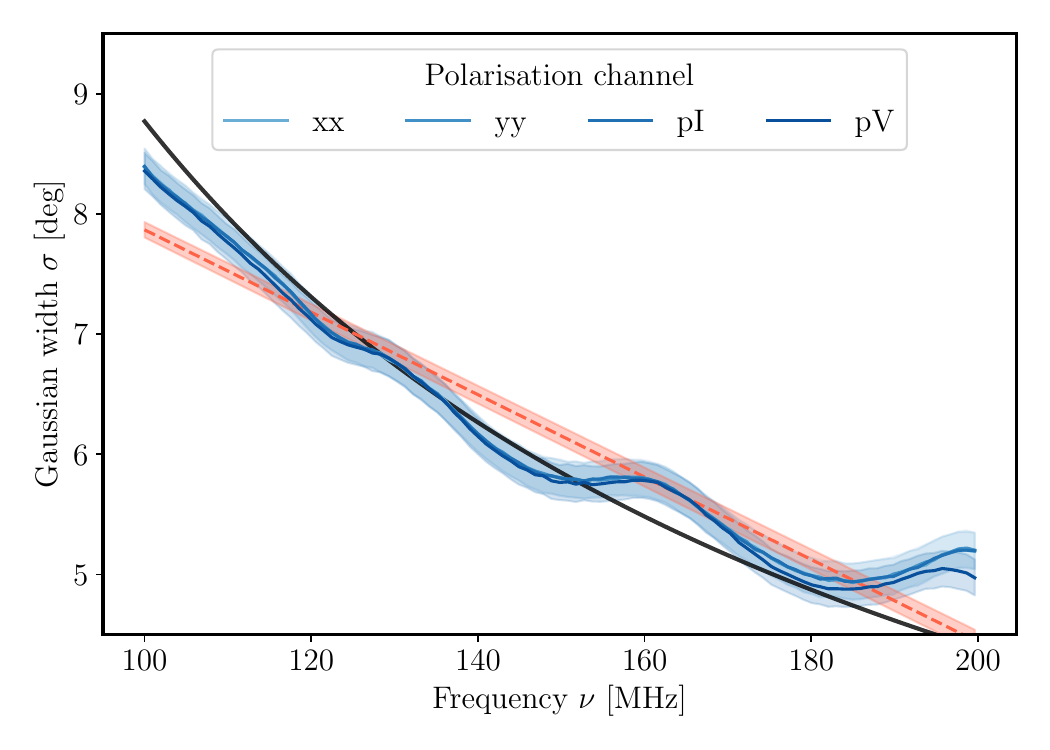}
    \caption{Full-width half-maximum of the Gaussian fit of the HERA beam, for different polarisation channels, as a function of frequency (solid lines). The four data sets are fitting by a linear evolution in frequency (eq.~\ref{eq:beam_fit}), resulting in the dashed line. Shaded areas represent the 68\% confidence intervals. The solid black line represents the evolution given in equation~\eqref{eq:beam_vs_freq}.}
    \label{fig:fit_Gaussian_regression}
\end{figure}

As frequency increases, the width $\sigma_b$ of the beam decreases, as illustrated in figure~\ref{fig:fit_Gaussian_regression}. This dependency is expected as we have the full-width half-maximum of the beam equal to
\begin{equation}
\label{eq:beam_vs_freq}
    \theta_\mathrm{FWHM} = \frac{\lambda(z)}{b},
\end{equation}
where $\lambda(z) \equiv 21\,\mathrm{cm} \times (1+z)$ is the redshifted 21\,cm wavelength and $b$ is the characteristic baseline length of the array. For HERA H1C IDR3 data, this value is equal to the mean baseline length, after applying inverse noise-variance weights: $b=19.58\,\mathrm{m}$. In figure~\ref{fig:fit_Gaussian_beam}, we show this evolution to compare with the Gaussian width fitted to the simulation. To model the frequency-response more precisely, we fit the data with a straight line to the obtained widths for four polarisation channels and obtain
\begin{equation}
\label{eq:beam_fit}
    \sigma_b(\nu)/[\mathrm{deg}] = - ( 0.0343 \pm 0.0003) \nu/\nu_0 + (11.30 \pm 0.04)
\end{equation}
for $\nu_0=1\,\mathrm{MHz}$. We compare the beam obtained with this linear model to the true beam at $\nu = 113.7~\mathrm{MHz}$ and for pI polarisation on figure~\ref{fig:fit_Gaussian_beam} and we find a reasonably good match.

\begin{figure}
    \centering
    \includegraphics[width=\columnwidth]{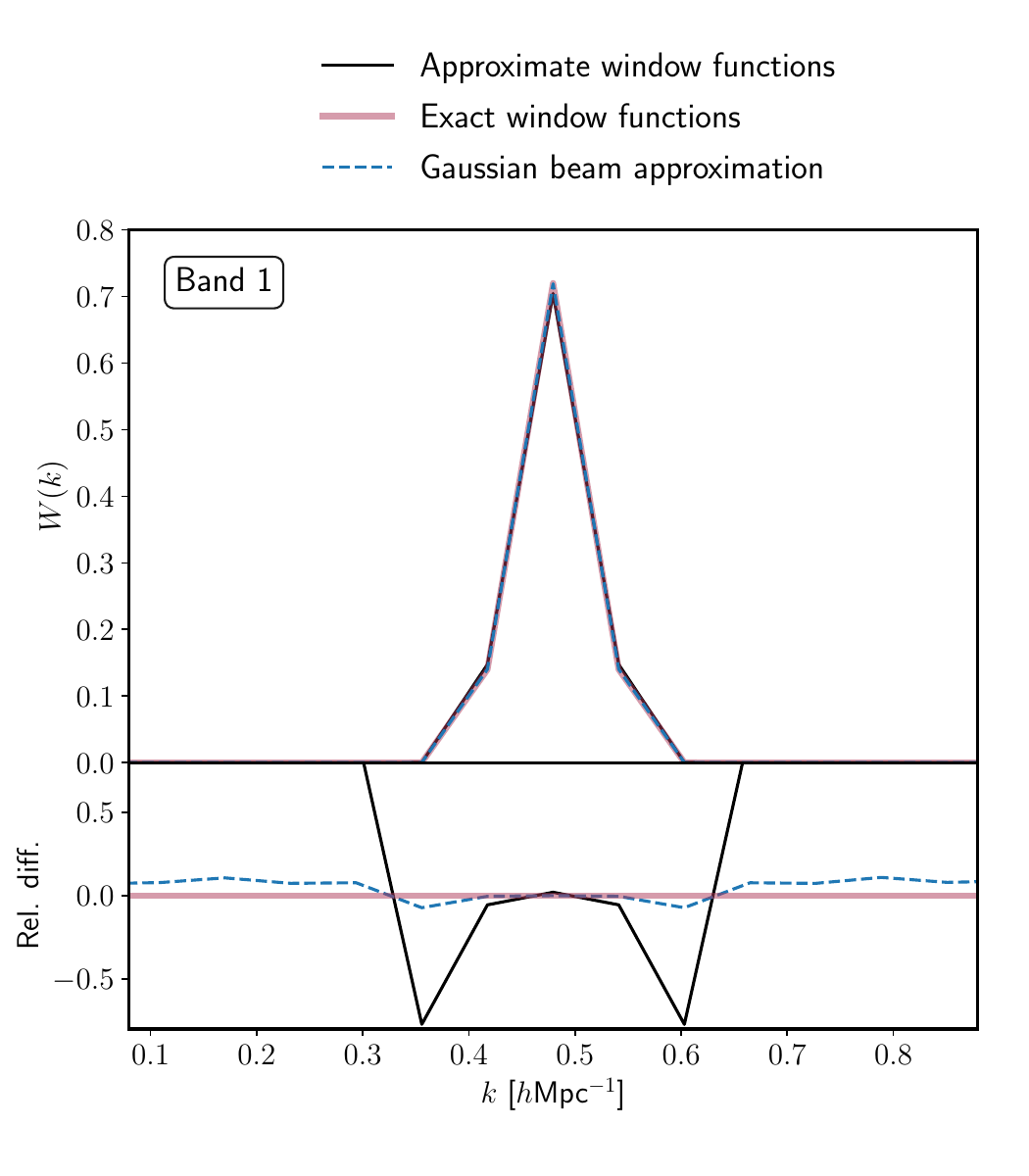}
    \caption{Relative difference between the exact spherical window functions and different approximations for Band~1 of the HERA analysis, at $k=0.48~h\mathrm{Mpc}^{-1}$.}
    \label{fig:rel_diff_approx_sph_wf}
\end{figure}

The resulting spherical window functions, once inverse-noise variance weights have been applied, are shown in figure~\ref{fig:spherical_wf}. We see that the Gaussian beam is a very good approximation of the exact window functions, despite the suppression of the tails of the beam seen in figure~\ref{fig:fit_Gaussian_beam}. A closer look shows that the tails of the window functions are underestimated by about $5\%$ in the Gaussian approximation -- this missing power is found in a slightly higher peak, compared to the approximate window functions underestimating the tails by as much as $77\%$, as illustrated in figure~\ref{fig:rel_diff_approx_sph_wf}.

\section{Investigating the outlier}
\label{app:outlier}

\begin{figure}
    \centering
    \includegraphics[width=\columnwidth]{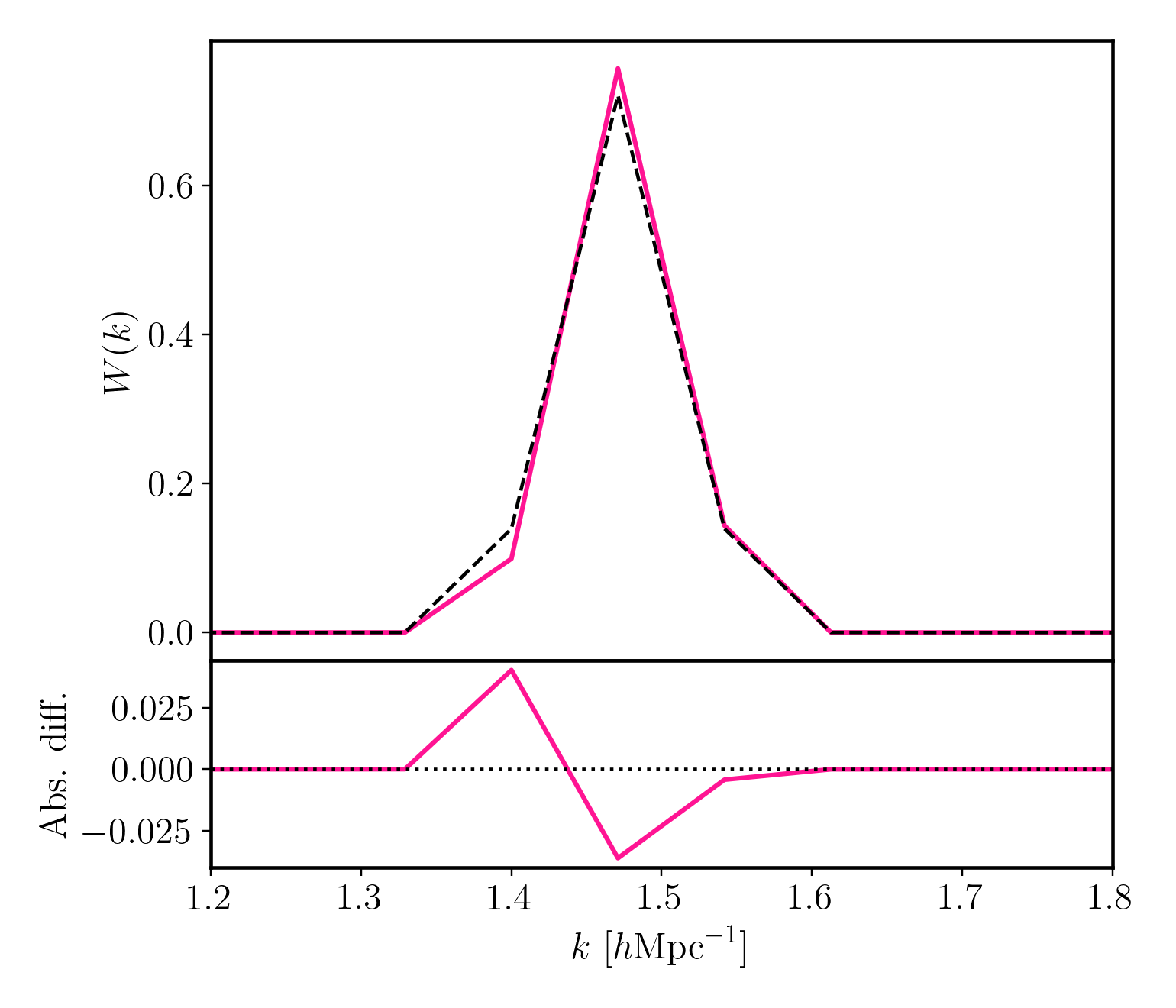}
    \caption{Difference between the outlier spherical window function (solid pink line) and a (shifted) regular neighbour (dashed black line).}
    \label{fig:diff_outlier}
\end{figure}

In this appendix, we investigate the amplitude difference observed in figure~\ref{fig:baselines_in_bins} between the spherical window function centred at $k=1.47\,h\mathrm{Mpc}^{-1}$ and its neighbours. As illustrated on figure~\ref{fig:diff_outlier}, we find that the larger amplitude can be explained by a weaker tail on the low-$k$ side, which is compensated by a larger amplitude when normalising.

\begin{figure}
    \centering
    \includegraphics[width=\columnwidth]{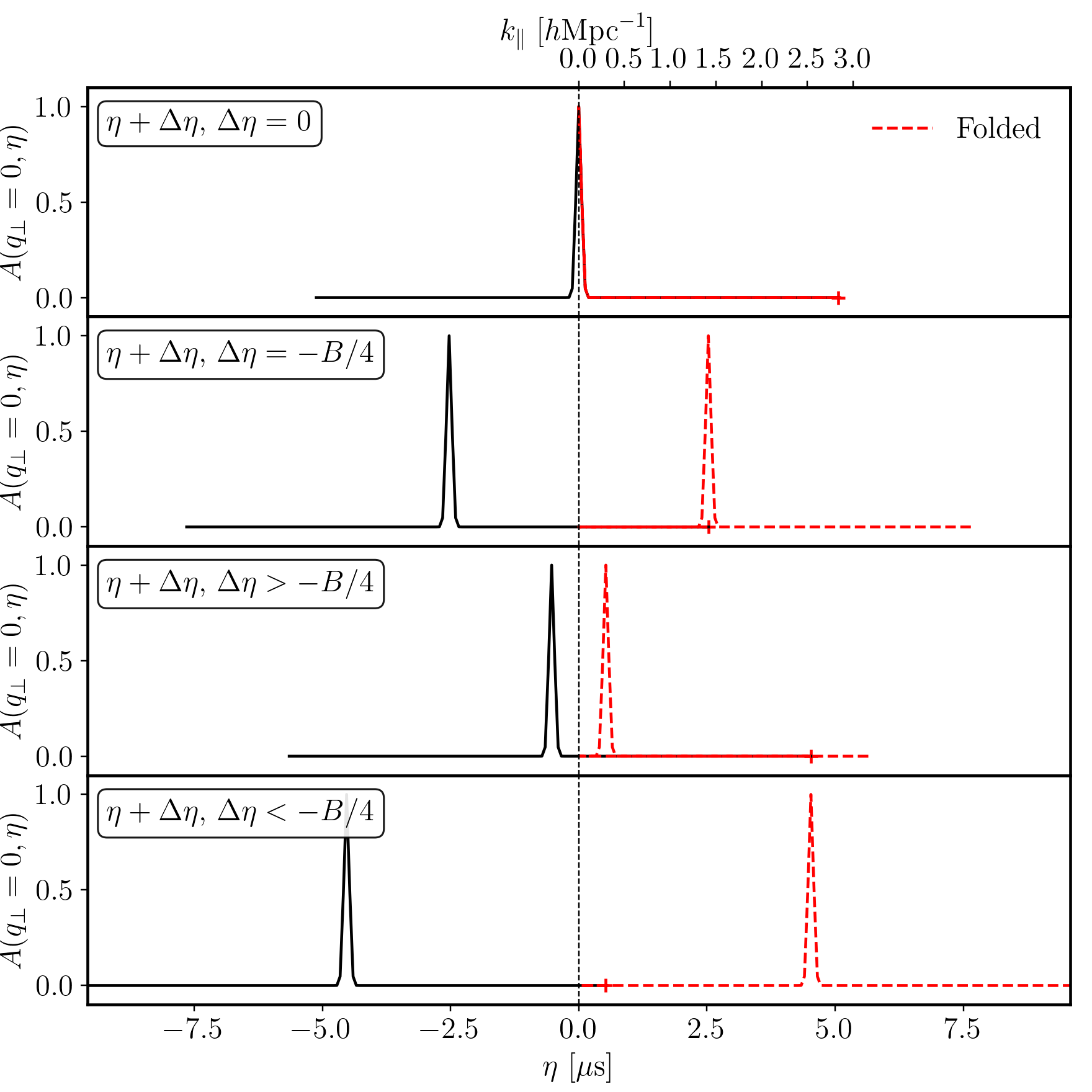}
    \caption{Normalised Fourier transform of the beam $\tilde{A}(\mathbf{q}_\perp, \eta)$ at $q_\perp = 0$, where the signal is maximal, as a function of $\eta$ and for different coordinate changes: folding ($\eta^\prime=\vert \eta \vert$) and translation ($\eta^\prime = \eta + \Delta \eta$).}
    \label{fig:outlier_geometry}
\end{figure}

This difference stems from the cylindrical binning of the window functions. Indeed, when building the spherical window functions, we subsequently bin the Fourier transform of the beam $\tilde{A}(\mathbf{q}_\perp, \eta)$ along $k_\perp$ and $k_\parallel$ using equations~\ref{eq:def_qperp} and \ref{eq:def_eta}.
In particular, the window function centred on $k=k_0$, mapping to $k_\parallel = k_{\parallel, 0}$ and $\tau=\tau_0$, receives contributions from all the modes along $\eta$ such that 
\begin{equation}
k_{\parallel,0}=\frac{2\pi}{\alpha(z)} \left| \eta + \tau_0 \right| = \frac{2\pi}{\alpha(z)} \tau_0,
\end{equation}
which is reached for $\eta = 0$ and $\eta=-2\tau$. The latter is only achieved if $\tau \leq B/4$ where $B$ is the bandwidth considered. Changing coordinates from $\eta$ to $k_\parallel$ is equivalent to translating and folding  $\tilde{A}(\mathbf{q}_\perp, \eta)$, illustrated on figure~\ref{fig:outlier_geometry}.
On this figure, we see that:
\begin{description}
    \item For small translations (third panel), $\vert \tau \vert <B/4$, two modes will contribute to each $k$-bin: $\eta = 0$ and $\eta=-2\tau$. Since $\eta=-2\tau$ is far from the centre of the beam, its contribution will always amount to zero and hence the total contribution (the mean of the two) will be $W(\eta=0)/2$.
    \item For large translations (fourth panel), $\vert \tau \vert>B/4$, only one mode, $\eta=0$, contributes to each $k$-bin. The difference with the above case is washed out by the normalisation.
    \item For a translation exactly equal to a quarter of the whole $\eta$ range (second panel), or $\vert\tau \vert = B/4$, the left-hand side of the window function ($k_\parallel < k_{\parallel,0}$) will be probed twice, with one of the contributions being zero, whilst the right-hand side of the window function will be probed only once (($k_\parallel > k_{\parallel,0}$) ), explaining the asymmetry in the resulting window function and the weaker tail at low $k$.
\end{description}

\bsp	
\label{lastpage}
\end{document}

%% file: author-list.tex
\author[A. Gorce et al.]{	Adélie  Gorce$^{1}$\thanks{Email: \href{mailto:adelie.gorce@mcgill.ca}{adelie.gorce@mcgill.ca}},
	Samskruthi  Ganjam$^{1}$,
	Adrian  Liu$^{1}$,
	Steven G. Murray$^{2}$,
	Zara  Abdurashidova$^{3}$,
	Tyrone  
\newauthor
	Adams$^{4}$,
	James E. Aguirre$^{5}$,
	Paul  Alexander$^{6}$,
	Zaki S. Ali$^{3}$,
	Rushelle  Baartman$^{4}$,
	Yanga  Balfour$^{4}$,
	Adam 
\newauthor
	P. Beardsley$^{2,7,\dagger}$,
	Gianni  Bernardi$^{8,9,4}$,
	Tashalee S. Billings$^{5}$,
	Judd D. Bowman$^{2}$,
	Richard F. Bradley$^{10}$,
\newauthor
	Philip  Bull$^{11,12}$,
	Jacob  Burba$^{13}$,
	Steven  Carey$^{6}$,
	Chris L. Carilli$^{14}$,
	Carina  Cheng$^{3}$,
	David R. DeBoer$^{15}$,
\newauthor
	Eloy  de~Lera~Acedo$^{6}$,
	Matt  Dexter$^{15}$,
	Joshua S. Dillon$^{3}$,
	Nico  Eksteen$^{4}$,
	John  Ely$^{6}$,
	Aaron  Ewall-Wice$^{3,16}$,
\newauthor
	Nicolas  Fagnoni$^{6}$,
	Randall  Fritz$^{4}$,
	Steven R. Furlanetto$^{17}$,
	Kingsley  Gale-Sides$^{6}$,
	Brian  Glendenning$^{18}$,
\newauthor
	Deepthi  Gorthi$^{3}$,
	Bradley  Greig$^{19}$,
	Jasper  Grobbelaar$^{4}$,
	Ziyaad  Halday$^{4}$,
	Bryna J. Hazelton$^{20,21}$,
\newauthor
    Jacqueline N. Hewitt$^{22,23}$,
	Jack  Hickish$^{15}$,
	Daniel C. Jacobs$^{2}$,
	Austin  Julius$^{4}$,
	MacCalvin  Kariseb$^{4}$,
\newauthor
	Nicholas S. Kern$^{3,23}$,
	Joshua  Kerrigan$^{13}$,
	Piyanat  Kittiwisit$^{12}$,
	Saul A. Kohn$^{5}$,
	Matthew  Kolopanis$^{2}$,
\newauthor
	Adam  Lanman$^{1}$,
	Paul  La~Plante$^{3,24,25}$,
	Anita  Loots$^{4}$,
	David Harold~Edward MacMahon$^{15}$,
	Lourence 
\newauthor
    Malan$^{4}$,
	Cresshim  Malgas$^{4}$,
	Keith  Malgas$^{4}$,
	Bradley  Marero$^{4}$,
	Zachary E. Martinot$^{5}$,
	Andrei  Mesinger$^{26}$,
\newauthor
	Mathakane  Molewa$^{4}$,
	Miguel F. Morales$^{20}$,
	Tshegofalang  Mosiane$^{4}$,
	Abraham R. Neben$^{23}$,
	Bojan
\newauthor
    Nikolic$^{6}$,
	Hans  Nuwegeld$^{4}$,
	Aaron R. Parsons$^{3}$,
	Nipanjana  Patra$^{3}$,
	Samantha  Pieterse$^{4}$,
	Jonathan C. 
\newauthor
    Pober$^{13}$,
	Nima  Razavi-Ghods$^{6}$,
	James  Robnett$^{14}$,
	Kathryn  Rosie$^{4}$,
	Peter  Sims$^{1}$,
	Hilton  Swarts$^{4}$,
\newauthor
	Nithyanandan  Thyagarajan$^{27,14}$,
	Pieter  van~Wyngaarden$^{4}$,
	Peter K.~G. Williams$^{28,29}$,
	Haoxuan  Zheng$^{23}$
\\
$^{1}$ Department of Physics and McGill Space Institute, McGill University, 3600 University Street, Montreal, QC H3A 2T8, Canada\\
$^{2}$ School of Earth and Space Exploration, Arizona State University, Tempe, AZ 85287-1404, USA\\
$^{3}$ Department of Astronomy, University of California, Berkeley, CA 94720-3411, USA\\
$^{4}$ South African Radio Astronomy Observatory, Black River Park, 2 Fir Street, Observatory, Cape Town, 7925, South Africa\\
$^{5}$ Department of Physics and Astronomy, University of Pennsylvania, Philadelphia, PA 19104-6396, USA\\
$^{6}$ Cavendish Astrophysics, University of Cambridge, Cambridge CB3 0HE, UK\\
$^{7}$ Department of Physics, Winona State University, Winona, MN 55987, USA\\
$^{\dagger}$ NSF Astronomy and Astrophysics Postdoctoral Fellow\\
$^{8}$ INAF-Istituto di Radioastronomia, via Gobetti 101, 40129 Bologna, Italy\\
$^{9}$ Department of Physics and Electronics, Rhodes University, PO Box 94, Grahamstown, 6140, South Africa\\
$^{10}$ National Radio Astronomy Observatory, Charlottesville, VA 22903, USA\\
$^{11}$ Jodrell Bank Centre for Astrophysics, University of Manchester, Manchester, M13 9PL, UK\\
$^{12}$ Department of Physics and Astronomy,  University of Western Cape, Cape Town, 7535, South Africa\\
$^{13}$ Department of Physics, Brown University, Providence, RI 02906, USA\\
$^{14}$ National Radio Astronomy Observatory, Socorro, NM 87801, USA\\
$^{15}$ Radio Astronomy Lab, University of California, Berkeley, CA 94720-3411, USA\\
$^{16}$ Department of Physics, University of California, Berkeley, CA 94720-3411, USA\\
$^{17}$ Department of Physics and Astronomy, University of California, Los Angeles, CA 90095-1547, USA\\
$^{18}$ National Radio Astronomy Observatory, Socorro, NM 87801-0387, USA\\
$^{19}$ School of Physics, University of Melbourne, Parkville, VIC 3010, Australia\\
$^{20}$ Department of Physics, University of Washington, Seattle, WA 98195-1560, USA\\
$^{21}$ eScience Institute, University of Washington, Seattle, WA 98195-1560, USA\\
$^{22}$ MIT Kavli Institute, Massachusetts Institute of Technology, Cambridge, MA 02139, USA\\
$^{23}$ Department of Physics, Massachusetts Institute of Technology, Cambridge, MA 02142, USA\\
$^{24}$ Department of Computer Science, University of Nevada, Las Vegas, NV 89154, USA\\
$^{25}$ Nevada Center for Astrophysics, University of Nevada, Las Vegas, NV 89154, USA\\
$^{26}$ Scuola Normale Superiore, 56126 Pisa, Italy\\
$^{27}$ Commonwealth Scientific and Industrial Research Organisation (CSIRO), Space \& Astronomy, P. O. Box 1130, Bentley, WA 6102, Australia\\
$^{28}$ Center for Astrophysics, Harvard \& Smithsonian, Cambridge, MA 02138, USA\\
$^{29}$ American Astronomical Society, Washington, DC 20006, USA
\vspace{3cm}
}